\documentclass[zamm,a4paper,fleqn]{w-art}
\usepackage{times}
\usepackage{w-thm}
\usepackage[]{graphicx}
\begin{document}
%%    The information for the title page will be placed between
%%    \begin{document} and \maketitle. The order of most entries
%%    is determined by the class file and can not be changed by
%%    rearranging them. The maketitle command follows after the
%%    abstract.
%%
%%    Most of the following commands will be completed by the publisher.
%%
%%    The copyrightyear is defined in the .clo file as the first argument
%%    of the copyrightinfo command. If the copyrightyear differs from that
%%    value it might be adjusted by the following definition:
%%
%% \renewcommand{\copyrightyear}{2003}% uncomment to change the copyrightyear.
%%
\DOIsuffix{theDOIsuffix}
%%
%% issueinfo for header and copyright line
\Volume{VV}
\Issue{I}
\Month{MM}
\Year{YYYY}
%%
%%    First and last pagenumber of the article. If the option
%%    'autolastpage' is set (default) the second argument may be left empty.
\pagespan{1}{}
%%
%%    Dates will be filled in by the publisher. The 'reviseddate' and
%%    'dateposted' (Published online) entry may be left empty.
\Receiveddate{XXXX}
\Reviseddate{XXXX}
\Accepteddate{XXXX}
\Dateposted{XXXX}

%% \pretitle{Editor's Choice}

%% We have a short and a long form for the title. The short form
%% (optional argument) goes into the running head.

\title[Short Title]{Magnetohydrodynamic experiments on cosmic magnetic fields}

%% Please do not enter footnotes or \inst{}-notes into the optional
%% argument of the author command. The optional argument will go into
%% the header.  If there is only one address the marker \inst{x} may be
%% omitted.

%% Information for the first autor.
\author[F. Stefani]{Frank Stefani\inst{1,}%
  \footnote{Corresponding author,~e-mail:~\textsf{F.Stefani@fzd.de}, 
            Phone: +49\,351\,260\,3069, 
            Fax: +49\,351\,260\,2007}}
\address[\inst{1}]{Forschungszentrum Dresden-Rossendorf, P.O. Box 510119 Dresden, Germany}
\author[A. Gailitis]{Agris Gailitis\inst{2}\footnote{gailitis@sal.lv}}
\address[\inst{2}]{Institute of Physics, University of Latvia, LV-2169 Salaspils 1,
Latvia}
%%
%%    Information for the third author
\author[G. Gerbeth]{Gunter Gerbeth\inst{1}\footnote{G.Gerbeth@fzd.de}}

%%
%%    Information for the second author
%\author[S. Author]{Second Author\inst{1,2,}\footnote{Second author footnote.}}
%\address[\inst{2}]{Second address}
%%
%%    Information for the third author
%\author[T. Author]{Third Author\inst{2,}\footnote{Third author footnote.}}
%%
%%    \dedicatory{This is a dedicatory.}
\begin{abstract}
It is widely known that cosmic magnetic fields, i.e. the                %!!!
fields of planets, stars, and galaxies, are produced by the
hydromagnetic dynamo effect in moving electrically conducting
fluids. It is less well known that cosmic magnetic fields play
also an active role in cosmic structure formation  by enabling
outward transport of angular momentum in accretion disks via the
magnetorotational instability (MRI). Considerable theoretical
and computational progress has been made in understanding both
processes. In addition to this, the last ten years have seen
tremendous efforts in studying both effects in 
liquid metal experiments.
In 1999, magnetic field self-excitation was observed in the
large scale liquid sodium facilities in Riga and Karlsruhe.
Recently, self-excitation was also obtained in the French
''von K\'{a}rm\'{a}n sodium''
(VKS) experiment. An MRI-like mode was found on the background of a
turbulent spherical Couette flow at the University of Maryland. 
Evidence for MRI as the first instability of an
hydrodynamically stable flow was obtained 
in the ''Potsdam Rossendorf Magnetic Instability Experiment''
(PROMISE). In this review,  the history of dynamo and MRI related
experiments is delineated, and some directions of future work 
are discussed.
\end{abstract}
%% maketitle must follow the abstract.
\maketitle                   % Produces the title.
%% If there is not enough space inside the running head
%% for all authors including the title you may provide
%% the leftmark in one of the following three forms:

%% \renewcommand{\leftmark}
%% {F. Author: A short title}

%% \renewcommand{\leftmark}
%% {F. Author and S. Author: A short title}

%% \renewcommand{\leftmark}
%% {F. Author et al.: A short title}

%% \tableofcontents  % Produces the table of contents.
\section{Once upon a time...}

Magnetism has been known for approximately 3000 years. There is some evidence
that a hematite bar, found close to Veracruz (now Mexico),
had served the Olmecs as a simple compass \cite{CARLSON}.
In any case, the Chinese have built, probably in the first century 
B.C.,
a compass  in the form of a lodestone spoon that was 
freely turnable on a polished
bronze plate \cite{NEEDHAM}.
The old Greek philosophers, starting with Thales of Miletus \cite{ARISTOTLE}, 
were well aware of the attracting forces of lodestone, and
the Roman philosopher Lucretius (95?-55 B.C.)
described its action in an atomistic way \cite{LUKREZ}: "First, 
stream there must from off the lode-stone seeds.
Innumerable, a very tide, which smites 
by blows that air asunder lying betwixt 
the stone and iron. And when is emptied out 
this space, and a large place between the two 
is made a void, forthwith the primal germs 
of iron, headlong slipping, fall conjoined 
into the vacuum, and the ring itself 
by reason thereof doth follow after and go 
thuswise with all its body."

\begin{vchfigure}[]
\begin{center}
\includegraphics[width=0.7\linewidth]{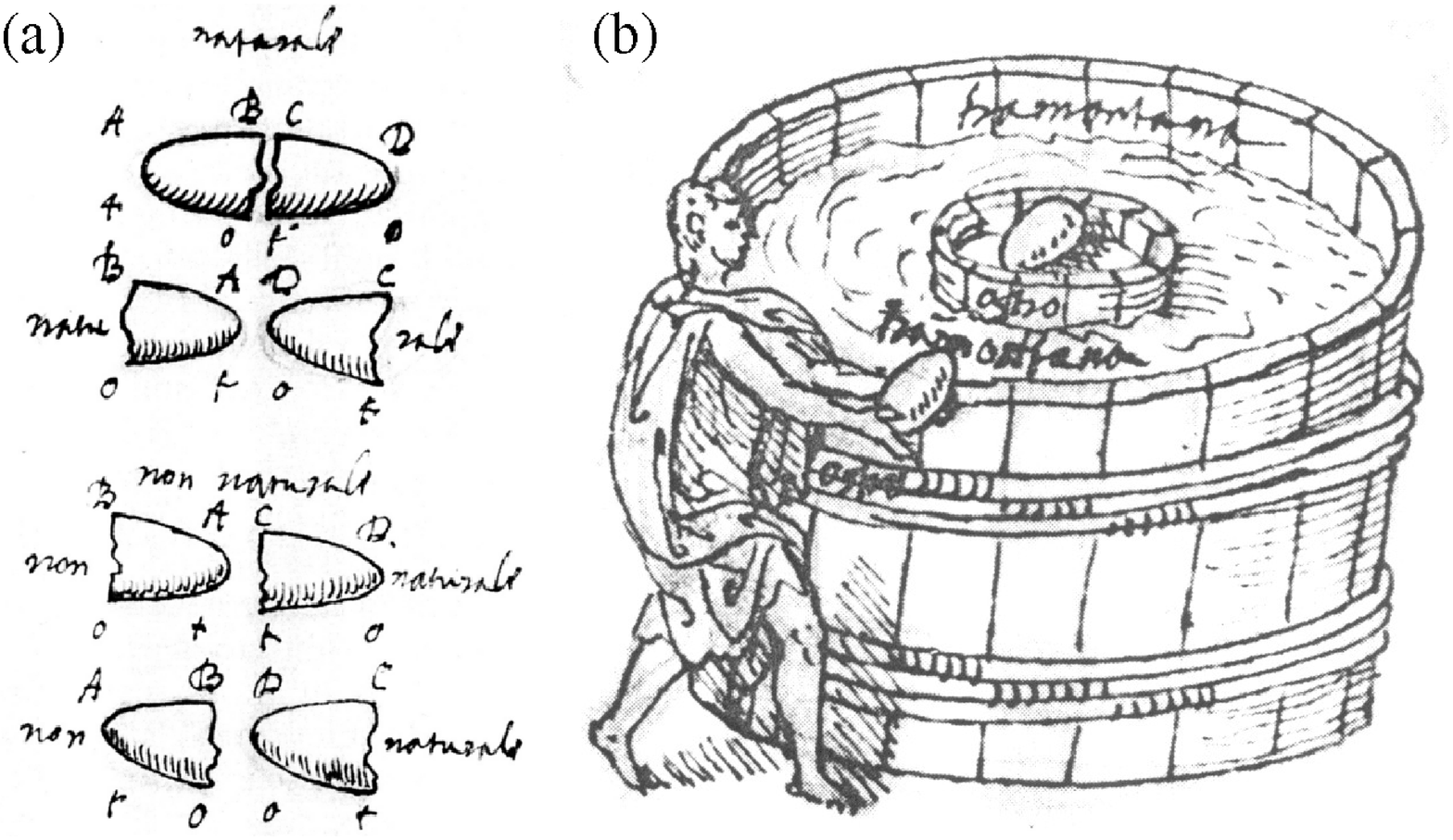}
\vchcaption{Lodestone experiments of Petrus Peregrinus. (a) Experiments showing the 
attracting and
repelling forces of a broken piece of loadstone. Upper configuration: "naturale", 
lower configuration: "non naturale". (b) A simple compass. 
Figures from \cite{PEREGRINUS}. }
\label{fig1}
\end{center}
\end{vchfigure}

As early as 1269, a first systematic experimental study of the 
attracting and repelling
forces of lodestone was published  by Petrus
Peregrinus in his
"Epistola de magnete" \cite{PEREGRINUS}.
For the first time, he defined the concept of polarity and  
distinguished the north and south poles of the magnet.
He was the first to formulate the law that poles of 
opposite polarity attract while poles 
of the same polarity repel each other (cf. Figure 1a).
Besides the construction of several compasses (cf. Figure 1b), he also
proposed a magnetic perpetuum mobile. 

Three centuries later, Peregrinus' work 
inspired William Gilbert to make 
his own experiments with small spheres of lodestone (``terrellae''), 
which led him, in 
1600, to the
conclusion that 
``...that the terrestrial globe is magnetic and is a loadstone 
\cite{GILBERT1600}''

However, this lodestone theory soon ran into trouble when 
the westward drift of the Earth's magnetic field declination 
was described by Gellibrand in  1635 \cite{GELLIBRAND},  and 
the detection of abrupt polarity reversals by David and
Brunhes in 1904/05 \cite{BRUNHES} has dealt it the 
ultimate deathblow.

Interestingly, it was not the well-studied magnetic field of the Earth, 
but the observation of magnetic fields in sunspots \cite{HALE}, 
that put Larmor on the right track speculating  \cite{LARMOR} that it could be
"...possible
for the internal cyclic motion to act after the
manner of the cycle of a
{\it self-exciting dynamo}, and maintain
a permanent magnetic field from insignificant
beginnings, at the expense of some of the energy
of the internal
circulation.''  This one-page communication, in which a natural 
process was explained in terms of a technical device \cite{SIEMENS,SIMONYI,WHEATSTONE},
was the birth certificate of the modern hydromagnetic 
theory of cosmic magnetic fields.

\section{Cosmic magnetism}
Wherever in the cosmos 
a large quantity of an electrically conducting fluid
is found in convection, one can also expect a 
magnetic field to be around.

The Earth is not the only planet in the solar system
with a magnetic field  \cite{MERRILL,STEVENSON2003}. 
Fields are produced inside the gas giants
Jupiter, Saturn and the ice giants
Uranus, and Neptune.  Possibly, a dynamo
had worked inside
Mars in the ancient past \cite{CONNERNEY}.
The Mariner 10 mission in 1974-75 had revealed the
magnetic field of Mercury \cite{NESS}
and there remain many puzzles
as to how it can be produced 
\cite{SOUTHWOOD,CHRISTENSENMERKUR,GLASSMEIERMERKUR,GLASSMEIERMERKUR2}.
The detection of the magnetic field of Ganymede, the
largest Jupiter moon,
was one of the major discoveries of
NASA's Galileo spacecraft mission in 1996 \cite{KIVELSON}.
The fact that Venus does not have a dynamo generated magnetic 
field has been attributed to the very slow rotation 
\cite{MERRILL}, but also 
to the stably stratified liquid core \cite{STEVENSON1983} of this planet.

The magnetic fields  of sunspots were discovered                  
by Hale (1908) at Mt. Wilson observatory,
thus proving evidence that natural magnetism is not a phenomenon
restricted to the Earth. With view on the tight relation of      
sunspots                                                      
and magnetic fields, sunspot observation turns into
a perfect test field for any theory of solar magnetism.
Still today, 
the 11 year periodicity of sunspots, their migration 
towards the
equator (the ''butterfly diagram''), and the occurrence of
grand minima which are superimposed upon the main periodicity
are the subject of intensive  investigations 
\cite{OSSENDRIJVER}.

Some main-sequence stars of spectral type A have remarkable
magnetic field strengths on the order of 1 T which are hardly explainable   %!!!
by dynamo action and which have been 
claimed to be remnants of the star's formation, i.e.  
''fossil fields'' \cite{BRAITHWAITE}.                                 %!!!
However, this magnetic field strength is rather 
moderate compared with that of other stars.
The field  of some
white dwarfs can easily reach values of 100 T, and 
even fields of 10$^{11}$ T have been 
ascribed to some anomalous
X-ray emitting pulsars \cite{KOUVELIOTOU}.

Large-scale magnetic fields of the order of 10$^{-9}$ T
are observed in many
spiral galaxies \cite{BECK}. Usually there is                      %!!!
a close correlation of the magnetic field structure with
the spiral pattern that indicates the relevance of                  %!!!
dynamo action, although by far not all problems 
with the origin and amplitude of 
galactic seed fields are solved 
\cite{GRASSORUBINSTEIN,KULSRUDZWEIBEL,DZIOURKIEVICH}.

Fascinating phenomena appear close to the centers of galaxies
which are usually occupied by supermassive black holes.
These are fed by so-called accretion disks \cite{BALBUSHAWLEYRMP}, 
a process 
which results 
typically in two oppositely directed jets of high-energetic 
particles that can fill vast volumes with magnetic field energy
\cite{KRONBERG}.
In our galaxy, these jets are rather weak but show 
a particularly interesting feature. Morris et al. recently  reported the
detection of a 
double-helix nebula in this outflow, not far from the galactic center,
which they described as an Alfv\'{e}n wave \cite{MORRIS}. 
However, we will see later
that this might well be connected with a typical dynamo action.

The working principle of accretion disks, which are the most 
efficient ''powerhouses'' in the universe 
\cite{POWERHOUSE} supplying energy for systems such as X-ray 
binaries, active galactic nuclei, and quasars through the 
release of gravitational potential energy, 
had been  a puzzle for 
a long time. The problem is that matter, before it can be accreted 
by the central object, has to get rid off its angular momentum. 
The molecular viscosity 
of such gas disks is much to small to explain the 
observed accretion rates of stars and black holes, so that  
turbulent
viscosity has to be assumed \cite{SHAKURASUNYAJEV}. The point is only:
why accretion disks are turbulent at all? Since they obey
Kepler's third law, i.e, 
their angular velocity decays as $r^{-3/2}$ with the radius, while the 
angular momentum increases as $r^{1/2}$, Rayleighs criterion must 
be applied stating
that rotating flows with radially increasing angular momentum 
are linearly stable for all Reynolds numbers \cite{RAYLEIGH}. 
In principle, the solution to this puzzle was already given 
in papers by Velikhov in 1959 \cite{VELIKHOV} and Chandrasekhar
in 1960 \cite{CHANDRA}, who had detected that 
a Taylor-Couette flow in the hydrodynamically stable regime could be
destabilized by an axially applied magnetic field. The astrophysical 
importance
of this ''magnetorotational instability'' was, however, noticed by
Balbus and Hawley in their seminal paper of 1991 \cite{BALBUSHAWLEY}.

Going beyond the galactic scale, we find  randomly tangled 
magnetic fields
also in galaxy clusters \cite{GOVONOFERETTI} which brings
us to the topic of {\it fluctuation dynamos} 
(or {\it small-scale dynamos})                  %!!!!
which have attracted much interest recently \cite{SCHEKOPLASMA,SCHEKOCHIHIN}.

\section{Some mathematical basics}

The temporal evolution of the velocity field $\bf{v}$ under the influence
of a magnetic field $\bf B$ is governed by the
Navier-Stokes equation
\begin{eqnarray}
\frac{\partial {{\bf{v}}}}{\partial t}+({\bf{v}}
\cdot \nabla) {\bf{v}}=
- \frac{\nabla p}{\rho} + \frac{1}{\mu_0 \rho}
(\nabla \times {\bf{B}})
\times {\bf{B}} + \nu \Delta  {\bf{v}}+{\bf{f}}_{d} \label{7} \; ,
\end{eqnarray}
where  $\rho$ and $\nu$ denote
the density and the kinematic viscosity of the fluid,
$p$ is the pressure, $\mu_0$ the magnetic permeability of the vacuum, 
and ${\bf{f}}_{d}$ symbolizes driving forces as, e.g.,                     %!!!
buoyancy in cosmic bodies or mechanical forcing by 
propellers in liquid metal experiments.
The magnetic field $\bf{B}$ in equation (1) is in general the  sum of an
externally applied magnetic field and the flow induced 
or self-excited magnetic field.

In order to derive the temporal evolution for $\bf B$ in a
fluid of electrical conductivity $\sigma$, we start
with Amp\`ere's law, Faraday's law, the divergence-free 
condition for the
magnetic field, and Ohm's law in moving conductors:
\begin{eqnarray}
\nabla \times {\bf{B}}= \mu_0 {\bf{j}} \\
\nabla \times {\bf{E}}= -\dot{{\bf{B}}} \\
\nabla \cdot {\bf{B}}=0 \\
{\bf{j}}=\sigma ({\bf{E}}+{\bf{v}} \times {\bf{B}}) \; .
\end{eqnarray}
Here, $\bf{E}$ denotes the electric field                      %!!!
and $\bf{j}$ denotes the electric current density.             %!!!
We have skipped the displacement current
in equation (2) as in most relevant cases 
the quasistationary approximation 
holds.
Taking the $curl$ of equations (2)  and (5),
inserting
equation (3), and assuming                                   %!!!
$\sigma$ to be constant in the considered                   %!!!
region, one readily arrives at the {\it induction equation}  %!!!
for the magnetic field:
\begin{eqnarray}
\frac{\partial {{\bf{B}}}}{\partial t}=\nabla
\times ({\bf{v}} \times {\bf{B}})
+\frac{1}{\mu_0 \sigma} \Delta {\bf{B}} \label{4} \; .
\end{eqnarray}
Obviously, the right hand side of equation (6) describes the 
competition between the
diffusion and the advection
of the field. For ${\bf v}=0$ equation (6) reduces to a 
vector heat equation and the field 
will decay within a typical 
time $t_d= {\mu_0 \sigma} l^2$,
with $l$ being a typical length scale of the considered system.        %!!!
Switching on the advection term, it can lead to an increase of
${\bf{B}}$ within a kinematic time $t_k=l/v$, where
$v$ is a typical velocity of the flow. If the
kinematic time
is smaller than the diffusion time, the net effect
can become positive, and the field will grow.
Comparing the diffusion time-scale with the kinematic time-scale
we get a dimensionless number that governs the "fate" of the
magnetic
field which is called magnetic Reynolds number $Rm$:
\begin{eqnarray}
Rm:=\mu_0 \sigma l v \; .
\end{eqnarray}
Depending on the flow pattern,                                  
the values of the critical $Rm$, at which the field starts 
to grow,
are usually in the range of 10$^1$...10$^3$. 
Most flows in cosmic bodies,
in which $Rm$ is large enough, will act as dynamos,
although there are a number of anti-dynamo theorems excluding 
too simple structures of the velocity field or the 
self-excited magnetic field \cite{COWLING,FEARN,ROBERTS94,IVERS88,KAISER2007}.

The competition between field dissipation and production
can also be understood in terms of
the energy balance.
Taking the scalar product of
the induction equation with ${\bf{B}}/\mu_0$,
and performing an integration by parts, we find for the                %!!!!
time evolution of the
magnetic energy
\begin{eqnarray}
\frac{d}{d t} \int_V \frac{{\bf{B}}^2}{2 \mu_0} dV=
-\int_V {\bf{v}} \cdot ({\bf{j}
\times {\bf{B}}})\; dV
-\int_V \frac{{\bf{j}}^2}{\sigma} dV 
-\frac{1}{\mu_0}\oint_S ({\bf E} \times {\bf B}) \cdot d\bf{S}
\label{5} \; .
\end{eqnarray}
In this form, the dynamo action can be interpreted in a
convenient  way: the time derivative of the magnetic field
energy equals the difference between the work  done (per time)
by the Lorentz forces on one side and the Ohmic and Poynting flux losses 
on the other side.
The Lorentz
force converts kinetic energy into magnetic energy, the Ohmic
dissipation converts magnetic energy into heat, the Poynting flux 
transports
electromagnetic energy across the surface $S$ of the considered 
volume $V$.

Besides the magnetic Reynolds number $Rm$, the coupled system of Eqs.
(1) and (6) is governed by some more dimensionless numbers: first
of all the well known Reynolds number $Re:=lv /\nu$, second 
the Hartmann number
$Ha:=B l \sqrt{\sigma /\nu \rho}$ which describes the square root 
of the ratio of magnetic to 
viscous forces. In some cases, the system behaviour is better described     %!!!
by the interaction parameter
(Stuart number) $N=\sigma B^2 l/(v \rho)=Ha^2/Re$
or the Lundquist number $S=:B l \sigma \sqrt{\mu_0/\rho}=Ha \sqrt{Pm}$,     %!!!
wherein
the magnetic Prandtl number is defined as the ratio of 
kinematic viscosity to magnetic diffusivity: $Pm:=\nu \mu_0 \sigma$.
Of course, more dimensionless numbers will enter the scene
when a particular forcing and/or global rotation of the system
is taken into account.

The coupled system of equations (1) and (6) can be treated with
varying complexity. For many technologically relevant cases, but also
for the ''helical MRI'' to be discussed later,
with $Rm<<1$, it will suffice to use the so-called inductionless 
approximation \cite{PRIEDE1}.  On the other extreme, 
one can study ''kinematic dynamo models'' by just solving equation (6) 
while supposing $\bf v$ to be fixed. In general, however, the treatment of 
most magnetic instabilities and 
of dynamically 
consistent dynamos 
requires the simultaneous solution of equations (1) and (6).

The numerical costs of the simulations are strongly 
governed by the relevant spatial dimension of the considered 
system. In some cases, including long cylindrical 
dynamos with axially invariant flows or long MRI experiments 
based on Taylor-Couette flows, 
it is appropriate to start with an analysis of normal modes
in axial and azimuthal directions giving an eigenvalue problem
in $r$ direction only. For dynamo or
MRI experiments in finite cylinders, 2D models in $r,z$ 
will be appropriate. Most expensive are, of course, 
fully coupled 3D simulations of Eqs. (1) and (6).

Dynamo relevant flows are in general turbulent, 
the question is only about the turbulence level and its role in the dynamo process. 
Commonly, one distinguishes 
between so-called {\it laminar} and  {\it mean-field} dynamo 
models. Laminar  models are described by the unchanged
equation (6) with neglected turbulence.
The self-excited magnetic field varies on
the same
length scale as the velocity field does.
Mean-field dynamo models, on the other hand, are relevant for
highly turbulent flows.
In this case the velocity and the magnetic field are
considered as superpositions of mean and fluctuating parts,
${\bf{v}}=\bar{\bf{v}}+{\bf{v}}'$ and
${\bf{B}}=\bar{\bf{B}}+{\bf{B}}'$. From equation (6) we             
get the
equation
for the mean part $\bar{\bf{B}}$,
\begin{eqnarray}
\frac{\partial {\bar{\bf{B}}}}{\partial t}=\nabla
\times (\bar{\bf{v}} \times \bar{\bf{B}}+{\bf{\cal{E}}})
+\frac{1}{\mu_0 \sigma} \Delta \bar{\bf{B}} \label{8} \; .
\end{eqnarray}
This equation for the mean field is identical
to equation (6) 
for the original field, except for one additional      
term
\begin{eqnarray}
{\bf{\cal{E}}}=\overline{{\bf{v}}' \times {\bf{B}}'} \; ,\label{9}
\end{eqnarray}
that
represents
the mean electromagnetic force (emf) due
to the fluctuations of the velocity and the magnetic
field.
The elaboration of mean-field dynamo models in the sixties by           
Steenbeck, Krause and R\"adler  \cite{STEENBECKKRAUSERAEDLER}
was a breakthrough
in dynamo theory (cf. \cite{KRRA80,RAEDLERRHEINHARDT}).
They had shown that, for homogeneous         %!!!
isotropic turbulence, the mean electromotive                       %!!!
force takes on the form                                    %!!!        
\begin{eqnarray}
{\bf{\cal{E}}}=\alpha \bar{\bf{B}} 
- \beta \nabla                                                           
\times \bar{\bf{B}} \; ,\label{10}                                      
\end{eqnarray}
with a
parameter $\alpha$ that is non-zero only for             %!!!
non-mirrorsymmetric velocity fluctuations ${\bf{v}}'$                           %!!!
(''cyclonic motion'' \cite{PARKER}) 
and a parameter $\beta$  that describes the enhancement                
of the electrical resistivity due to turbulence. The fact
that helical fluid motion can induce
an emf that is {\it parallel} to the magnetic field             
is now commonly known as the $\alpha$-effect.
Dynamo models based on the  $\alpha$-effect have played an
enormous role in the
study of solar and galactic magnetic fields, and we will
later explain the physics of the Karls\-ruhe experiment
in terms of a mean-field model with the $\alpha$-effect.           

A promising way to combine the credibility of direct numerical    %!!!
simulations with the convenience (and robustness) of                   %!!!
mean-field models is to carry out                                      %!!!
global 3D simulations at affordable spatial resolution, and to         %!!!   
extract then the mean-field coefficients by means of a test-field method %!!! 
\cite{SCHRINNER1,SCHRINNER2,GIESECKE,BRANDENBURGRAEDLERSCHRINNER}.   %!!!

\section{Why doing liquid metal experiments?}

During recent decades tremendous 
progress has been made                
in the analytical understanding and the  
numerical treatment of flows with high $Rm$, including dynamos, which
has been reported in
dozens of monographs and review articles 
\cite{BUSSE78,MOFFATT78,KRRA80,INGLIS,
ROBERTSJENSEN,ROBERTSSOWARD,CHILDRESSGILBERT,HOLLERBACH96,MERRILL,
FEARN98,ROBERTSGLATZMAIER2000,BUSSE2000,BUSSE2002,RUEHOLLBUCH,
BRANDENBURGSUBRAMANIAN}.

As for the geodynamo, to take one example, recent numerical simulations
\cite{GLATZMAIERROBERTS,KAGEYAMA,KUANGBLOXHAM,BUSSEGROTETILGNER,CHRISENSENOLSENGLATZMAIER,
CHRISTENSENAUBERT,WICHTOLSON,STELLMACH,HARDERHANSEN,
AUBERTAURNOUWICHT,STELLMACH2008}
share their main results with features of the Earth's magnetic
field,
including the dominance of the axial dipolar component, weak
non-dipolar structures, and, in some cases, full polarity
reversals,
a behaviour that is well known from paleomagnetic
measurements (for a recent overview, see \cite{KONOROBERTS}).

Despite those successes, a number of unsolved problems          
remain. The simulations of the Earth's dynamo, to remain in this
picture for the moment,
are carried out in parameter regions far from the real one.     
This concerns, in particular, the Ekman number $E$
(the ratio of the rotation
time scale to the viscous time scale)  and the magnetic
Prandtl number $Pm$ (the ratio of the magnetic diffusion       
time to the                                                     
viscous diffusion time). The Ekman number
of the Earth is of the order 10$^{-15}$, the magnetic
Prandtl number is of the order 10$^{-6}$. Present
numerical
simulations are carried out for values 
as small as                                                 
$E \sim 10^{-5}$
and $Pm \sim 0.1$. The                                         
wide gap between real and numerically tractable
parameters is, of course, a continuing source of               
uncertainty about the physical reliability of those
simulations. The usual way in fluid dynamics to deal with
parameter discrepancies of this sort, namely to apply
sophisticated
turbulence models, is presently hampered by the lack of
validated turbulence models for fluids that are both fast
rotating and strongly interacting with a magnetic field.
Here is the crucial point where laboratory
experiments are unavoidable in order 
to collect knowledge about the turbulence structure in the
(rotating or not) dynamo regime.

With this critical attitude towards simulations, one must likewise
admit that none of the real cosmic bodies can be put into
a {\it Bonsai form} to be studied in laboratory. Taking again
the geodynamo
as a (striking) example,
it is not possible to actualize all of the
dimensionless numbers in an equivalent experimental set-up. 
A liquid sodium experiment of 1 m radius 
would have to rotate with 10$^8$ (!) rotations per second in order to
reach the Ekman number of the Earth, which amounts to twice 
the speed of light 
at the rim of the vessel.

So what, then, can we actually learn from liquid metal experiments ?

{\it First}, it is worthwhile to 
verify experimentally that hydromagnetic dynamos work at all. 
In theory and numerics, kinematic dynamo action has been proved for a 
large variety of more or less smooth velocity fields or pre-described 
distributions of turbulence parameters.  However, liquid metal flows,
at the necessary $Rm$, will be highly turbulent. 
Further, most dynamo simulations have been
carried out in spherical geometry. What happens  
when we are using cylindrical vessels instead of spheres?
How important is the correct implementation of the non-local 
boundary conditions for the magnetic field, which is trivial for 
spherical geometry 
but requires sophisticated methods in other, e.g. cylindrical, geometry       %!!!
\cite{STEFANIRAEDLER,JCP,PHYSREVE,ISKAKOV,GUERMOND,GISSINGER,GIESECKE2008}?
Later, when discussing the VKS dynamo, we will see 
that even slight modifications of the experimental design 
can teach a lot about the role of turbulence, boundary conditions,
and the distribution of different dynamo sources.

{\it Second}, if one was lucky to make a hydromagnetic 
dynamo running, how can the exponential field growth 
be stopped, how does the  dynamo saturate?
Roughly speaking, dynamo saturation is nothing than
an application of Lenz's rule stating that an induced current acts 
against the source of its own generation.
How this saturation works in detail, depends strongly on the
mechanical constraints the flow is experiencing.
All of the  present laboratory dynamos
comprise mechanical installations to drive and
guide the flow (propellers, guiding blades, etc.).
Obviously, the fewer installations are present in the
fluid, the more freedom has the flow to be modified and 
re-organized by the Lorentz forces. It would be most interesting
to drive the
flow purely
by convection, as in the Earth's outer core. However, it seems to be  
impossible
to reach velocities sufficient for dynamo action in a purely           
convective way in
laboratory experiments, as discussed, e.g. in \cite{TILGNER2000}.
Hence, all present laboratory experiments have to find a
compromise between
a mechanical forcing of the flow and the degree of
freedom of the flow for
the magnetic field back-reaction.

This brings us to the {\it third} point: Besides its influence on the 
large scale flow, the 
magnetic field back-reaction may also change the turbulence properties of
the flow. Sometimes this effect is considered the most 
important one that dynamo experiments may help to understand, as they
provide an interesting
test-case for MHD turbulence models (a summary of the latter 
can be found in \cite{VERMA}). Those models, once validated, could  
gain  reliability
when applied to such hard problems as magnetic field generation
in the Earth's core
. 
But ''turbulence model validation'' sounds much easier 
as it is in reality. Even the
simple flow measurement in liquid sodium is a problem
in its own right, let alone the measurements of all sorts 
of correlation functions which might be important for
the validation of turbulence models.

A {\it fourth} topic, which is intimately connected with 
the issue of turbulence modification, is 
the destabilizing  role magnetic fields can have on flows.
Typically dynamo experiments and experiments on the  magnetorotational 
instability are of a similar size, making it worth to hunt for
new instabilities in the presence of 
(self-excited or externally applied)
magnetic fields. In addition to this, there is a large variety
of wave phenomena to be studied in rotating magnetized flows.

A {\it fifth} issue that could possibly 
be addressed by dynamo experiments
has to do with the distinction between steady and oscillatory
dynamo states. Typically, these transitions occur at so-called
exceptional points of the spectrum of the non-selfadjoint
dynamo operator, and polarity reversals have been
described as noise triggered relaxation oscillations in the vicinity of
such points \cite{REVERSALPRL,EPSL,GAFD}. However, reversals
can occur for a wide variety of bistable systems \cite{HOYNG},
and experiments can be helpful to distinguish between 
different reversal scenarios.

\section{The experiments in detail}

In the following we will concentrate on the most 
important experimental efforts related to the understanding of
the origin and the action of 
cosmic magnetic fields. We will start with the four experiments 
that have already shown homogeneous dynamo action, and then move 
to experiments devoted to wave phenomena and 
magnetic instabilities in liquid metals. For the sake of 
shortness, we have to skip some very interesting older experiments 
like those of Lehnert \cite{LEHNERT} (cf. \cite{LEHNERTMOMOMO} 
for a very amuzing account of these experiments in Stockholm)
and Gans \cite{GANS}, but also the impressive series of 
liquid metal experiments on Alfv\'{e}n waves which have been 
summarized by Gekelmann \cite{GEKELMAN}. Another topic omitted 
is the search for self-excitation phenomena in fast breeder 
reactors \cite{BEVIR,PIERSON,KIRKO,ALEMANY,PLUNIAN}, although 
this close connection was occasionally used as a political argument
to motivate dynamo experiments
(see \cite{STEENBECKLETTER}).  Slightly focusing on 
some newer experimental activities, we advise the reader 
to consult some former reviews on earlier experiments
\cite{ROBERTSJENSEN,BUSSE2000,LATHROP2001,RMP,SURV,COLGATEAN,
TILGNER2000,FAUVEPETRELIS,PETRELIS,MOMOMO}.

For decades, hydromagnetic dynamo experiments
seemed to be at the edge of technical
feasibility. The problem to achieve self-excitation is
that values of the critical $Rm$
for different flow geometries are of the order of 100.
For the best liquid metal conductor, sodium,
the product of conductivity and
magnetic permeability is approximately 10 s/m$^2$. Hence, to
get an $Rm$ of 100, the product of length and
velocity has to be 10 m$^2$/s. 
To reach this value one should have more than 1 m$^3$ sodium       
and use at least 100 kW of mechanical power to move it. Another possibility 
is of course to increase $Rm$  by simply using materials with a high
magnetic permeability. This brings us directly to the first
experiment on homogeneous, though not hydromagnetic, dynamo action.

\subsection{The dynamo experiments of Lowes and Wilkinson}

In the sixties of the 20 century, Lowes and Wilkinson have carried out a long-term
series of homogeneous dynamo experiments \cite{LOWI1,LOWI2,WILKINSON} at the 
University of Newcastle upon Tyne. 
The main idea of their experiments was already laid down in a 1958 paper by 
Herzenberg \cite{HERZENBERG} who had given the first rigorous existence
prove for a homogeneous dynamo consisting of two rotating small spheres
embedded in a large sphere (Figure 2a). Thus motivated, Lowes and Wilkinson 
started with the first homogeneous
dynamo using two rotating cylinders in a ``house-shaped'' surrounding
conductor (Figure 2b). The key point for the success of 
this and the
following  experiments
was the utilization of various ferromagnetic materials (perminvar, mild steel, 
electrical iron) 
making the magnetic Reynolds number large, simply by
a high relative magnetic permeability $\mu_r$ (between 150 and 250).

\begin{vchfigure}[h]
\begin{center}
\includegraphics[width=0.8\linewidth]{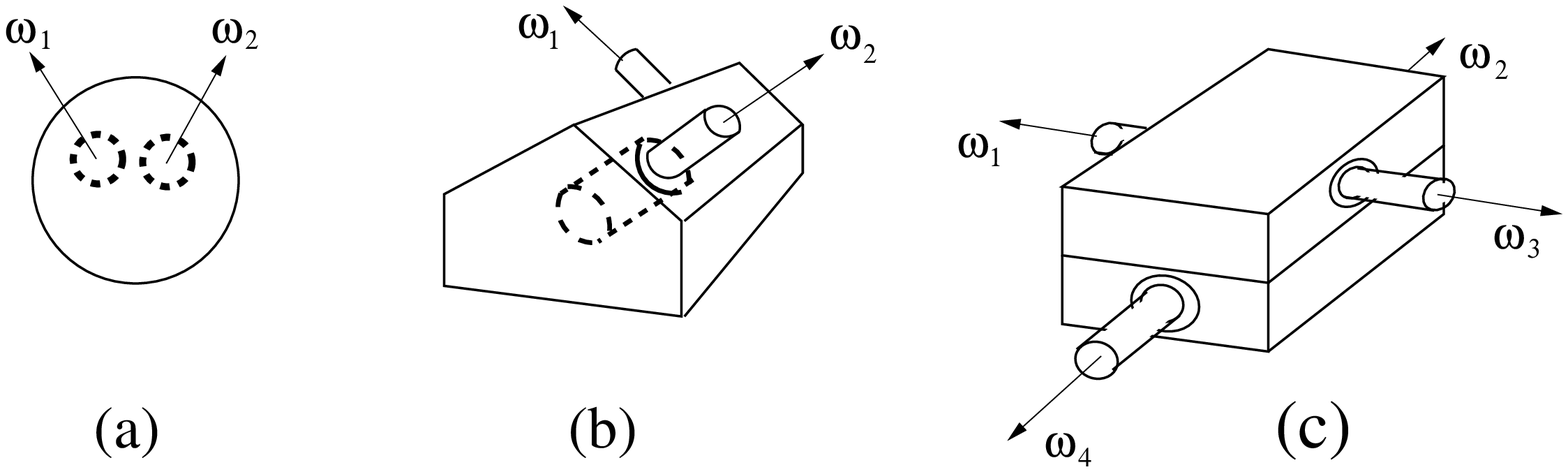}
\caption{(a) The Herzenberg dynamo. Two spheres rotate
around non-parallel
axes. (b) The first dynamo of Lowes and
Wilkinson. Two cylinders rotate in a ``house-shaped'' block. (c) The 
fifth  dynamo with four independent rotors.}
\end{center}
\end{vchfigure}

The history of these experiments is impressive, not only for         
their step-by-step improvements but also for the continuing        
comparison of the resulting field with geomagnetic features \cite{WILKINSON}.
Starting with a simple geometry of the rotating cylinders (Figure 2b), which
produced steady and oscillating magnetic fields, the design was
made more sophisticated (Figure 2c) so that finally 
it permitted the observation          
of field reversals . That way it was shown that a complex field
structure and
behaviour can result from comparatively simple                  
patterns of motion.

Needless to say, the experiments were flawed by the                          
use of ferromagnetic materials  and the  
nonlinear field behaviour which is inevitably connected with these            
materials. One attempt to get self-excitation with rotating 
non-magnetic
copper cylinders failed. And, although
homogeneous,  these dynamos were not suited to study the         
nontrivial
back-reaction of the magnetic field on the fluid motion, and
there was no chance to learn something about MHD turbulence.

\subsection{The dynamo experiments in Riga}

There is a long tradition at the 
Institute of Physics Riga, Latvia, 
to carry out dynamo related experiments.

\begin{figure}[h]
\begin{minipage}{72mm}
\begin{center}
\includegraphics[width=0.9\linewidth]{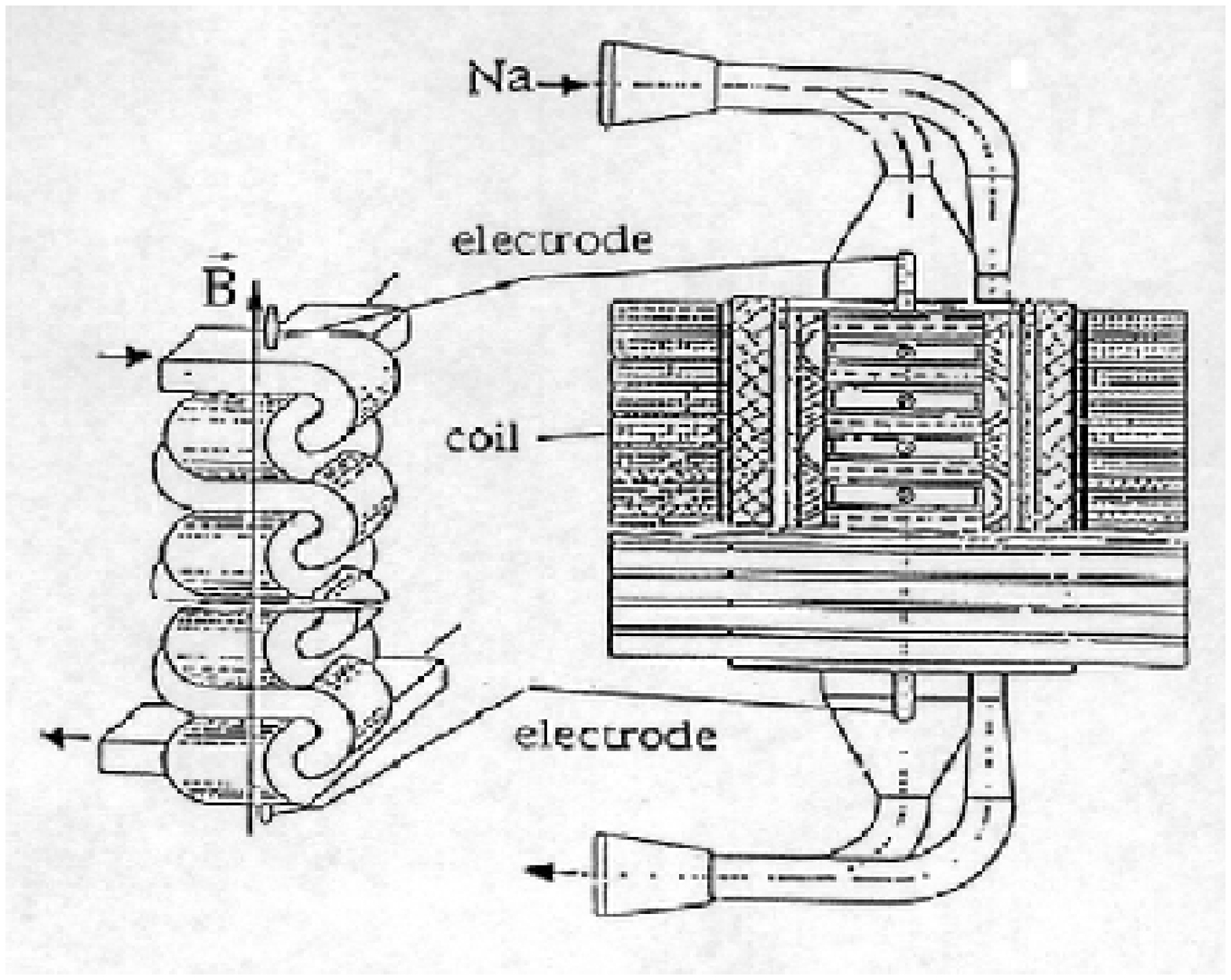}
\caption{The ``$\alpha$-box,'' the first dynamo-related experiment  
in Riga.
The sodium flow through the
helically interlaced channels produces an emf {\it parallel to the       
applied magnetic field.}}
\end{center}
\end{minipage}
\hfill
\begin{minipage}{72mm}
\begin{center}
\includegraphics[width=0.9\linewidth]{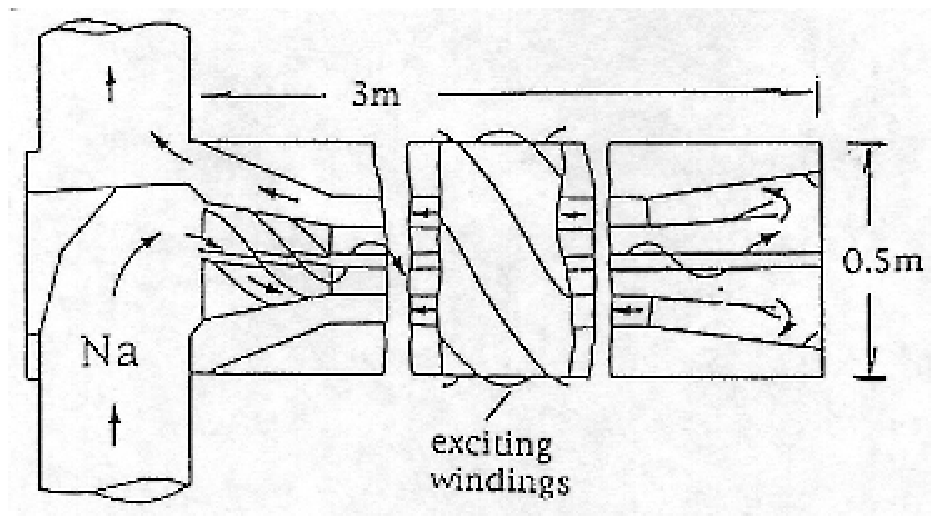}
\caption{The dynamo module of the 1987 experiment. Significant 
amplifications of externally applied magnetic fields
were measured, before the experiment had to be stopped due
to mechanical vibrations.}
\end{center}
\end{minipage}
\end{figure}

The first one, actually proposed by Max Steenbeck, was
intended to prove experimentally the $\alpha$ effect, i.e.
the induction of an electromotive force parallel to an 
applied magnetic field. This experiment, the  ``$\alpha$-box'' 
(Figure 3), consisted of two orthogonally 
interlaced copper channels through which sodium was pumped
with velocities up to 11 m/s. Interestingly, although  the very 
flow helicity 
${\bf v}\cdot (\nabla \times {\bf v})$ is zero everywhere in this set-up,
an $\alpha$ effect results from the non-mirrorsymmetry of the flow.
The main result of this experiment was that the induced voltage
between the electrodes (cf. Figure 3)
is proportional to $v^2$, i.e., it is 
independent of the flow direction, and that it reverses if the
applied magnetic field is reversed. The $\alpha$-effect
was therefore validated \cite{STEENBECKALPHA}. Interestingly, the
induced current was shown to increase slower than linearly 
with the applied magnetic field, a result which points to some 
quenching of $\alpha$ with increasing interaction parameter.

\begin{vchfigure}
\begin{center}
\includegraphics[width=0.8\linewidth]{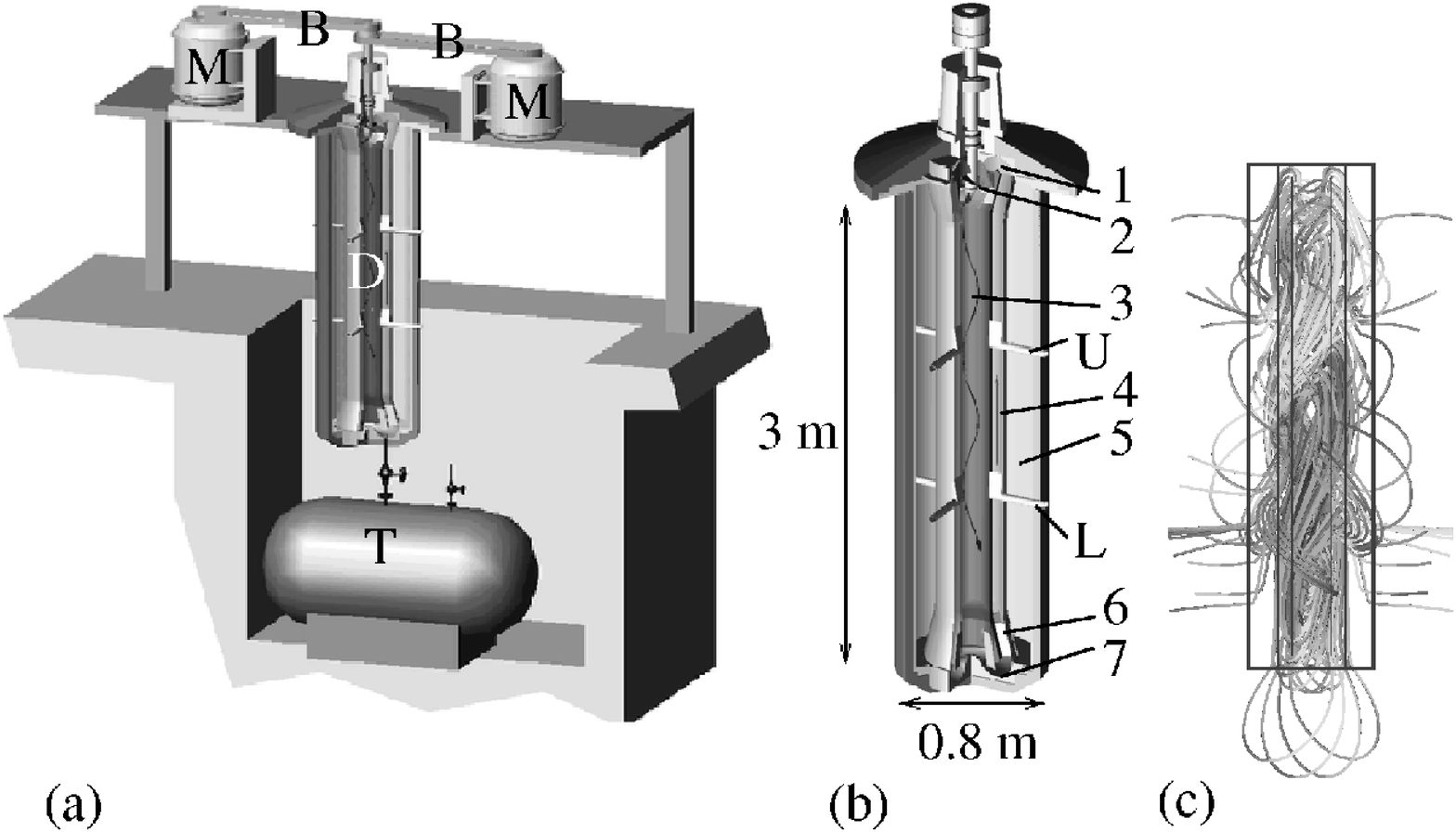}
\vchcaption{The Riga dynamo experiment and its eigenfield. (a) 
Sketch of the facility. M - Motors. B - Belts. D - Central dynamo module. T - Sodium tank.
(b) Sketch of the central module. 1 - Guiding blades. 
2 - Propeller.
3 - Helical flow region without any flow-guides,
flow rotation is maintained by inertia only.            
4 - Back-flow region. 5 - Sodium at rest. 6 - Guiding blades.
7 - Flow bending region. (c) Simulated magnetic eigenfield. The gray scale
indicates the vertical component of the field.}
\end{center}
\end{vchfigure}

A second experiment was also prepared 
in Riga but actually carried out in St. Petersburg
in 1986 \cite{STPETERSBURG} (Figure 4).
The principle idea of this, as well as of the later Riga dynamo 
experiment, traces back to Ponomarenko  \cite{PON} who had proved that
a helically moving, electrically conducting  cylinder
embedded in an infinite stationary conductor can show dynamo action.
This simple configuration was  analyzed in more
detail  by Gailitis and Freibergs \cite{GAFR76} who found a remarkably low
critical magnetic Reynolds number of 17.7 for the
convective instability.
By adding a back-flow, this convective instability can be
transformed into an absolute instability \cite{GAFR80}.

All this early numerical work, including the optimization \cite{GAI96}
of the main geometric
relations which led to the design of the Riga dynamo (Figure 5a,b), was
done with a one-dimensional eigenvalue solver.
For refined kinematic simulations a two-dimensional finite
difference code (in radial and axial direction)
was written whose  main advantage  is the
possibility to treat velocity structures
varying in axial direction, which is
indeed of relevance for the Riga
dynamo \cite{PLASMA}. The magnetic field structure as it
comes out of this code is illustrated in Figure 5c.

Much effort has been spent to fine-tune the
whole facility. The first step was to optimize the
main geometric relations, in particular the relations of the
three radii to each other and to the length of the system
 \cite{GAI96}.
The resulting shape of the central module of the dynamo
is shown in Figure 5b.
In a water dummy facility at the Dresden Technical University, 
many tests
have been carried out to optimize the velocity profiles         
\cite{CHRISTEN} and
to ensure the mechanical integrity of the system.                
All the experimental preparations were accompanied by
extensive numerical simulations. One main result of these simulations
was the optimization of the velocity profile with regard to
the limited motor power resources of around 200 kW. For helicity 
maximizing profiles (''Bessel function profiles'')
a critical $Rm$ as low as 12.0 (for the convective instability) and
14.7 (for the absolute instability) has been found, while the corresponding
numbers for the
measured (as far as possible optimized) 
profiles were 14.3 and 17.6, respectively 
\cite{KLUWER}.
Another result was the
prediction of the main features of the expected magnetic field,
i.e., its growth rate, frequency, and spatial structure, and the    
dependence of these features                                        
on the rotation rate of the propeller.                              

\begin{vchfigure}
\begin{center}                               
\includegraphics[width=0.8\linewidth]{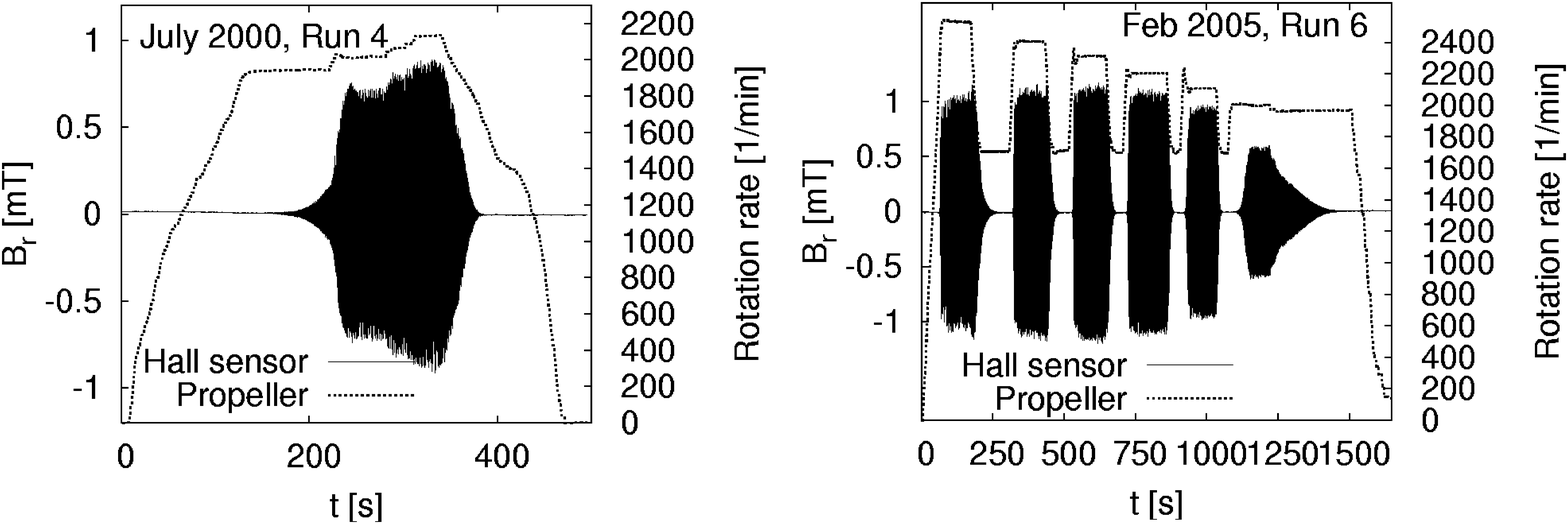}
\caption{Two experimental runs carried out in July 2000 and
in February 2005. 
Rotation rate of the motors, and
magnetic field measured at one Hall external sensor
plotted vs. time.
After the exponential increase of the magnetic field in the 
kinematic
dynamo regime, the dependence of the field level on the
rotation rate has been studied in the saturation regime.}                                              
\end{center}
\end{vchfigure}

At the present facility,
eight experimental campaigns have been carried out
between November 1999 and July 2007. In the  first campaign in November 1999,
a self-exciting field was documented for the first time in a liquid metal
dynamo experiment, although the saturated regime could not
be reached at that time \cite{PRL1}.
This had to be left until the July 2000 experiment \cite{PRL2}.
In June 2002, the radial dependence of the magnetic field
was determined by the use of Hall sensors and induction coils
situated on ''lances'' going
throughout the whole dynamo module.
In February and June 2003, first attempts were
made to measure the Lorentz force induced
motion in the outermost cylinder.
A novelty of the May 2004 campaign was the measurements
of pressure in the inner channel by a piezoelectric sensor
that was flash mounted at the innermost wall.
In February/March 2005, a newly developed permanent magnet probe
was inserted into the innermost cylinder in order to get
information about the velocity there, and two traversing rails with
induction coils  and Hall sensors were installed to
get continues field information along the
$z$-axis and across the whole diameter of the dynamo.
In July 2007, a newly 
developed magnetic coupler 
was installed to replace the outworn 
gliding ring seal.
More details about these results can be found in Refs.
\cite{RMP,PLASMA,MAHYD3,MAHYD1,MAHYD2,SURV}, and 
will also be published elsewhere.

In Figure 6 we document  two experimental runs carried out in 
July 2000 and in February 2005. It is clearly visible
that the magnetic field switches on and off when a critical value of
the propeller rotation rate is
crossed from below or above, respectively.

\begin{vchfigure}[h]
\begin{center}
\includegraphics[width=0.8\linewidth]{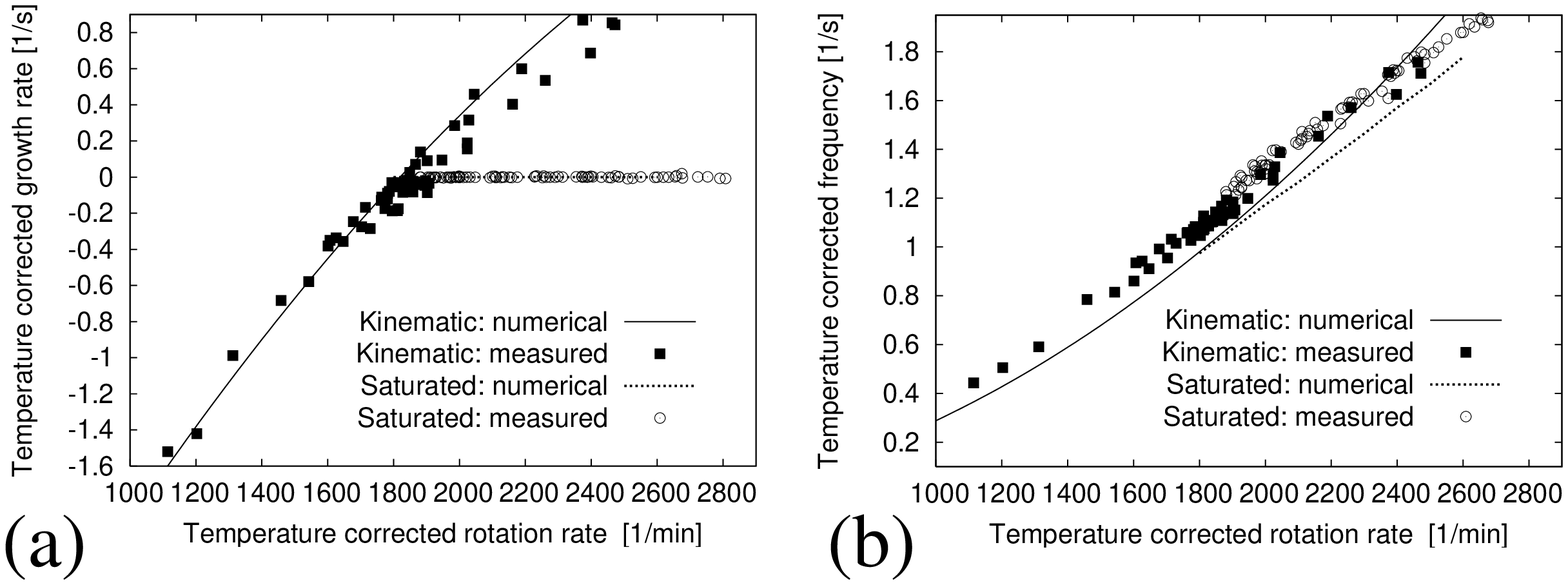}
\caption{Measured growth rates and frequencies of the magnetic eigenfield
for different
rotation rates  $\Omega$ in the kinematic and the saturation regime,
compared with the numerical
predictions.  $\Omega$, $p$, and $f$
at the temperature $T$ were scaled to
$(\Omega_{s}, p_s, f_s)=\sigma(T)/\sigma(157^{\circ}C) \,
(\Omega(T), p(T), f(T))$
as required by the scaling properties of equation (6).}
\end{center}
\end{vchfigure}

In  Figure 7 the temperature corrected measurement data for the growth rate (Fig
7a) and
the frequency (Fig. 7b) are shown in comparison with the corresponding
numerical results. The numerical curves in the kinematic regime were obtained 
with the
2D solver \cite{KLUWER,PLASMA} and were slightly corrected by the effect 
of the lower
conductivity of the stainless steel walls
that was estimated  separately  by a 1D solver.

As for the saturation regime we have modelled the most important
back-reaction effect within a simple one-dimensional model \cite{MAHYD3,PLASMA}.
This relies on the fact that, while the axial velocity component has to 
be rather constant from top to bottom
due to mass conservation, the azimuthal component $v_{\phi}$ can be easily
braked by the
Lorentz forces without any significant pressure increase.
In the inviscid
approximation, and considering only the $m=0$ mode of the
Lorentz force, we end up with the  ordinary differential
equation for the Lorentz force induced perturbation  $\delta v_{\phi}(r,z)$
of the azimuthal velocity component:
\begin{eqnarray}
\bar{v}_z(r,z) \frac{\partial}{\partial z} \delta v_{\phi}(r,z) =
\frac{1}{\mu_0 \rho} [(\nabla \times {\bf{B}}) \times
{\bf{B}}]_{\phi}(r,z)  \;\; .
\label{eq2}
\end{eqnarray}
This equation is now solved simultaneously with the induction equation,
both in the innermost channel where it describes
the downward braking of $v_{\phi}$, and in the back-flow channel where it
describes the upward acceleration of $v_{\phi}$. Both effects together lead to a 
reduction of the differential rotation and hence to a deterioration of the dynamo
capability of the flow.
The validity of this
self-consistent back-reaction model, which gives automatically a zero
growth rate, can be judged from the dependence of the
resulting eigenfrequency in Figure 7b. Actually, we see a quite reasonable 
correspondence
with the measured data, in particular with respect to the slope of the curve.

Only recently, a sophisticated  T-RANS (transient Reynolds average Navier-Stokes equation) 
model of the Riga dynamo experiment
has been developed at the Delft Technical University
\cite{KENJERES1,KENJERES2,KENJERES3}. 
This model, which incorporates the state of the 
art of hydrodynamic turbulence modelling under the influence of 
magnetic fields, has basically confirmed, and slightly 
improved, the main predictions of our simple one-dimensional
back-reaction model.

\subsection{The Karlsruhe dynamo experiment}

Historically it is interesting that not only 
the basic idea and the geophysical motivation, but also a
final
formula for the critical flow-rates for a sort of
Karlsruhe experiment can already be found
in a paper of 1967 \cite{GAI67}.
The idea was to substitute
real helical (''gyrotropic'') turbulence  by ''pseudo-turbulence''
actualized by
a large (but finite) number of parallel
channels with a helical flow inside.
Later, in 1975 , Busse considered a similar kind of 
dynamo \cite{BUSSE75} which prompted him to initiate the Karlsruhe dynamo
experiment which was then designed and carried out 
by R. Stieglitz and U. M\"uller.

In 1972, Roberts had proved  dynamo action for a
velocity pattern periodic in $x$ and $y$
that comprises both  a rotational flow and an
axial flow \cite{ROBERTSFLOW}.
The $\alpha$-part of the electromotive force for this flow  type
can be written in the form ${\bf{\cal{E}}}=-\alpha_{\perp}
(\bar{\bf{B}}-({\bf{e}}_z \cdot \bar{\bf{B}}) {\bf{e}}_z)$,
which represents an extremely anisotropic $\alpha$-effect that
produces only electromotive
forces in the $x$- and $y$-directions, but not
in the $z$-direction \cite{RAEDLER}.

In the specific realization of the Karlsruhe experiment (Figure 8), the
Roberts flow in each cell is replaced by a flow through two
concentric channels. In the central channel
the flow is straight, in the outer channel it is  forced by 
a "spiral staircase" on
a helical path (Figure 9).
This design principle of the Karlsruhe dynamo being given,
a fine tuning of the geometric relations was carried out
with the aim
to achieve a maximum $\alpha$ effect for a given power
of the pumps.
Such an optimization led to a number of 52 spin generators,
a radius of 0.85 m and a height of 0.7 m for the dynamo module.

\begin{figure}
\begin{minipage}{100mm}
\begin{center}
\includegraphics[width=0.9\linewidth]{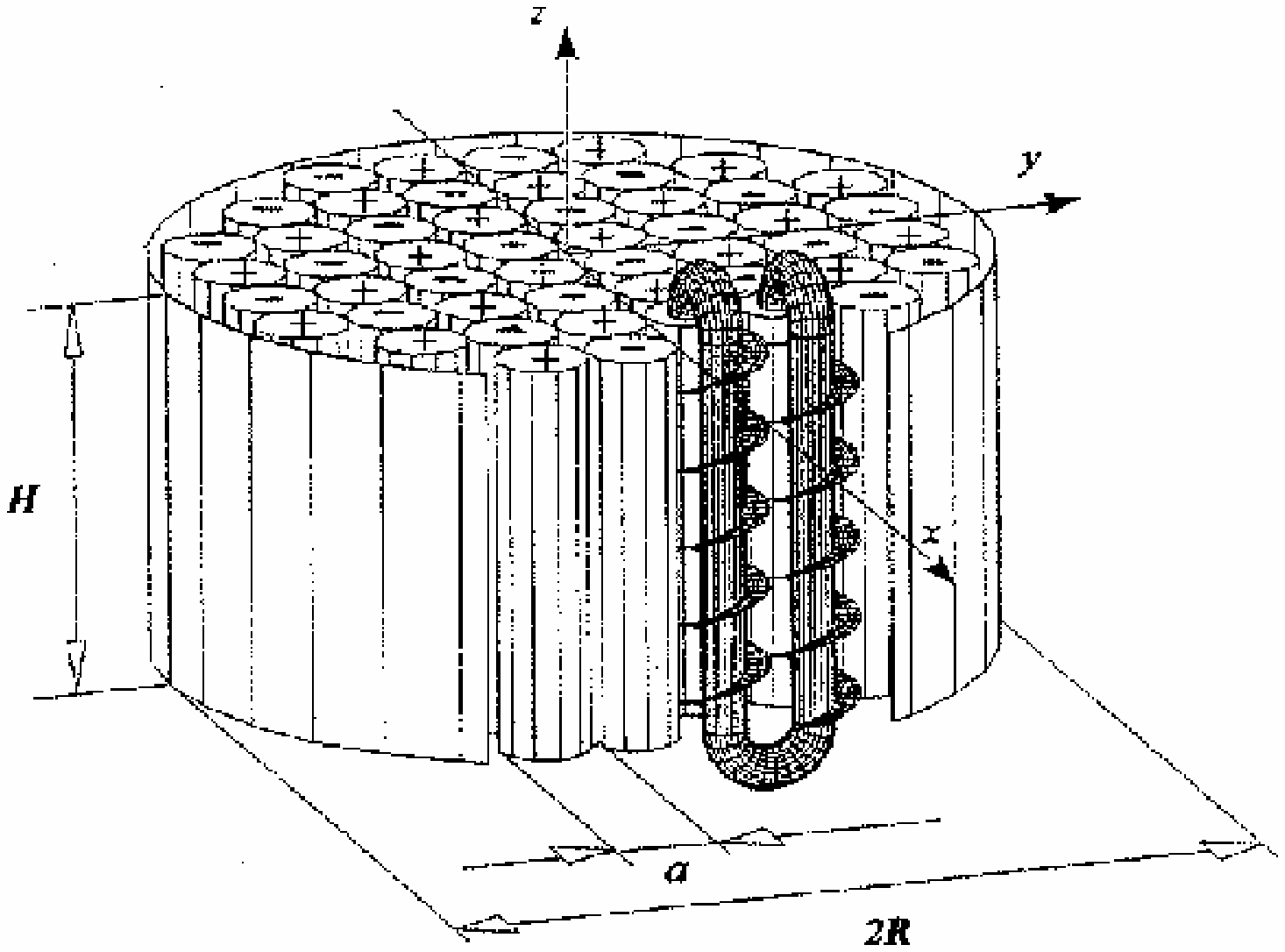}
\caption{Central part of the Karlsruhe dynamo facility.
The module consists of 52 spin-generators, each containing
a central tube with non-rotating flow and an outer tube where 
the flow is forced on a helical path. 
Figure courtesy of R. Stieglitz.}               
\end{center}
\end{minipage}
\hfill
\begin{minipage}{55mm}
\includegraphics[width=0.9\linewidth]{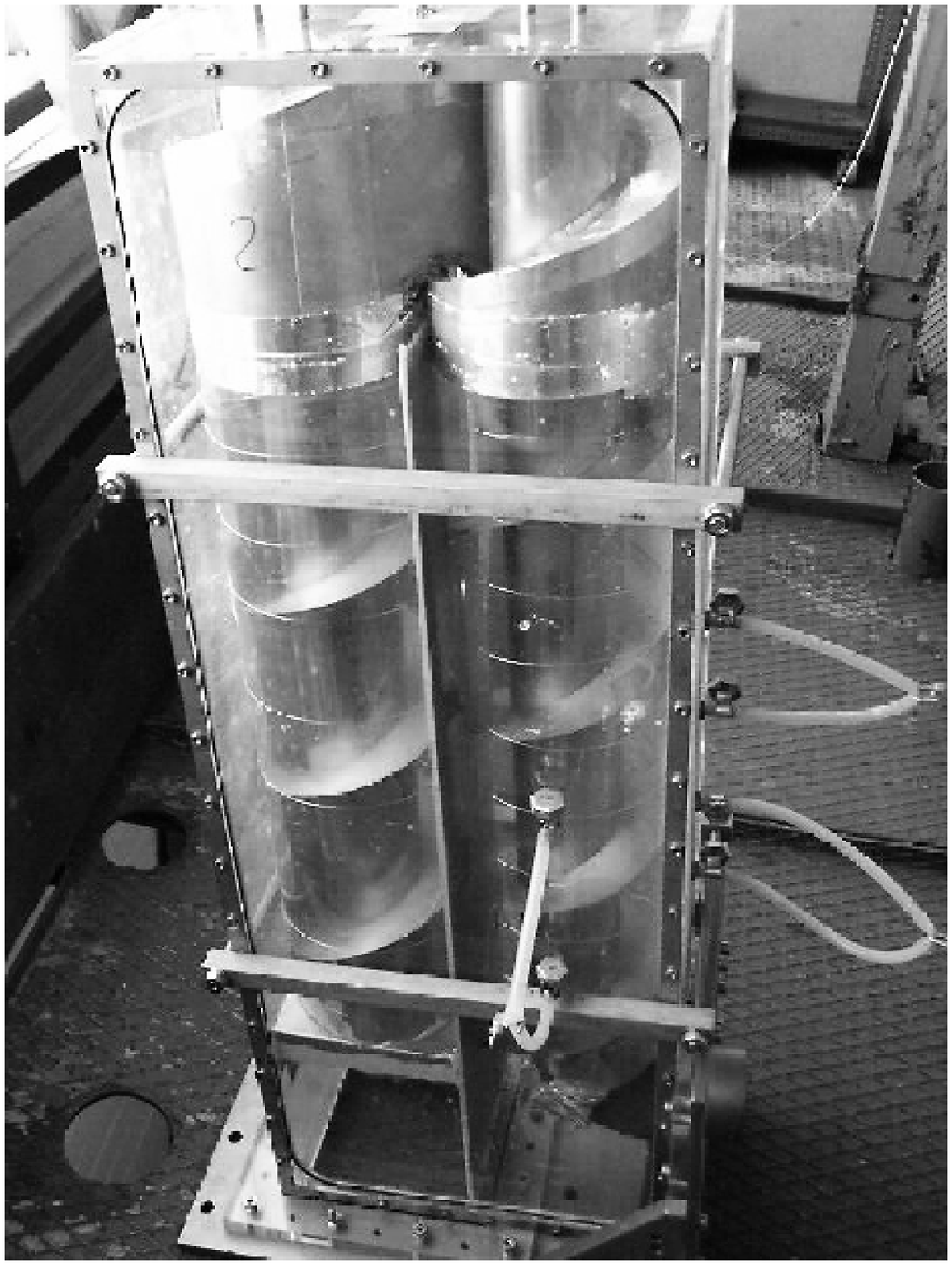}
\caption{A model of a spin-generator.
Figure courtesy of R. Stieglitz and Th. Gundrum.} 
\end{minipage}
\end{figure}

Figure 10 documents 
the experiment carried out in December 1999  \cite{MUST,STMU}. The scheme in         %!!!
Figure 10a depicts again the central dynamo module with the 52 
spin-generators,
and defines the coordinate
system for the location and direction of the Hall probes.
Figure 10b shows the measured $x$-component of the magnetic 
field. This signal was recorded  after the central flow rate
$\dot{V}_{C}$
was set to a constant value of 115 m$^3$/h and the flow rate
$\dot{V}_{H}$                                                          
in the helical ducts was increased from 95 m$^3$/h to
107 m$^3$/h at a time 30 s from the start of the
experiment. After approximately 120 s the field
starts
to saturate at a approximately 7 mT.

For the experimentally interesting region, the
isolines of the quantity $C=\mu_0 \sigma \alpha_{\perp} R$, which is a 
dimensionless     
measure of the $\alpha$-effect, and the experimentally obtained curves, 
are plotted in Figure 11.
The experimentally determined neutral line, separating dynamo and          
non-dynamo regions, corresponds to values                                  
of $C^{crit}$ in the region of 8.4...9.3. Hence, the numerical                
prediction, $C^{crit}=8.12$, resulting from mean-field theory \cite{RAEDLERMAHYD1},
was  quite reasonable.

\begin{figure}[h]
\begin{minipage}{90mm}
\begin{center}
\includegraphics[width=0.9\linewidth]{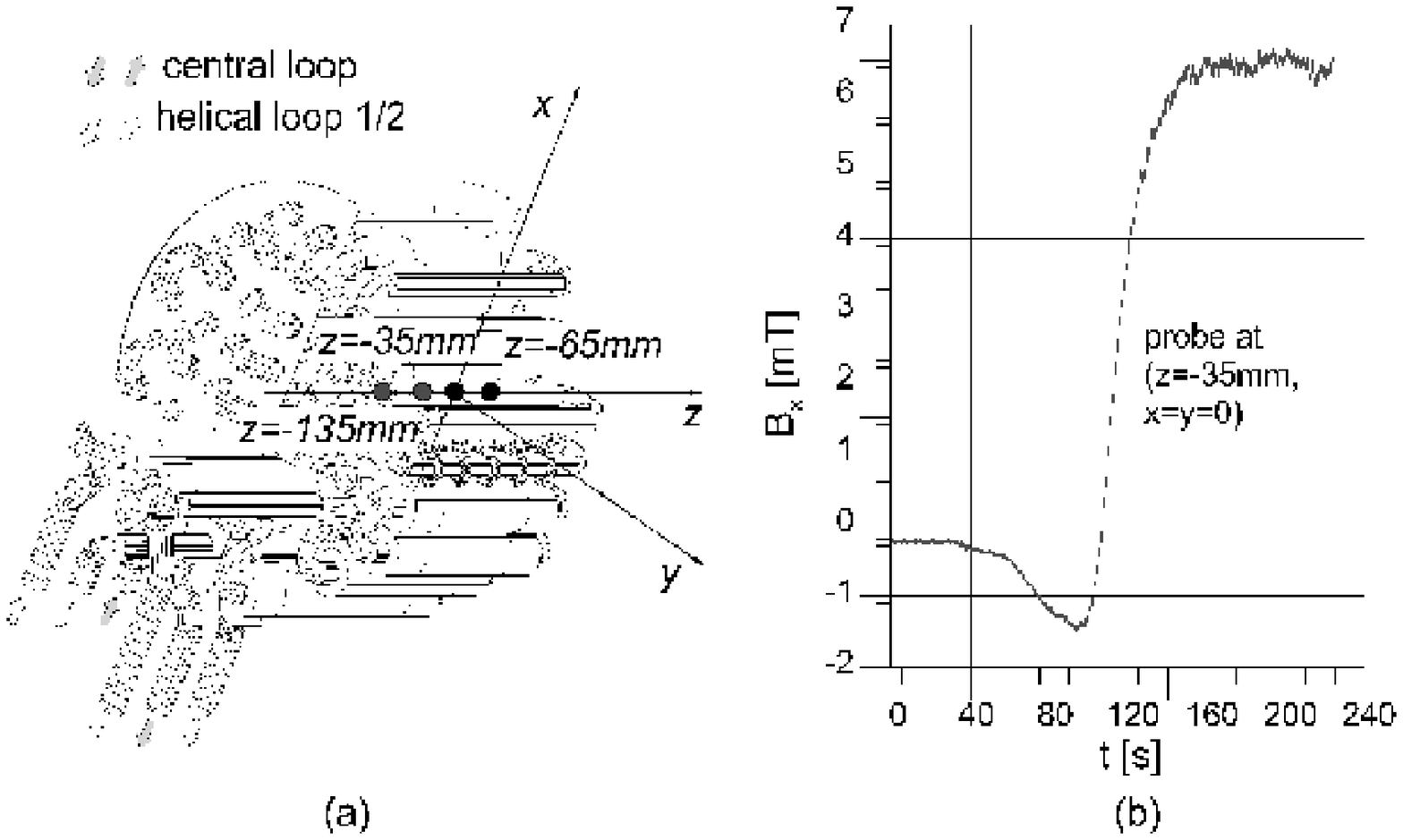}
\caption{Self-excitation and saturation in the Karlsruhe dynamo experiment.
(a) The dynamo module with the connections between the spin-generators and    
the supply pipes. (b)  Hall sensor signals of $B_x$
in the inner bore of the module. Figure courtesy of R. Stieglitz.}
\end{center}
\end{minipage}
\hfill
\begin{minipage}{60mm}
\begin{center}
\includegraphics[width=0.9\linewidth]{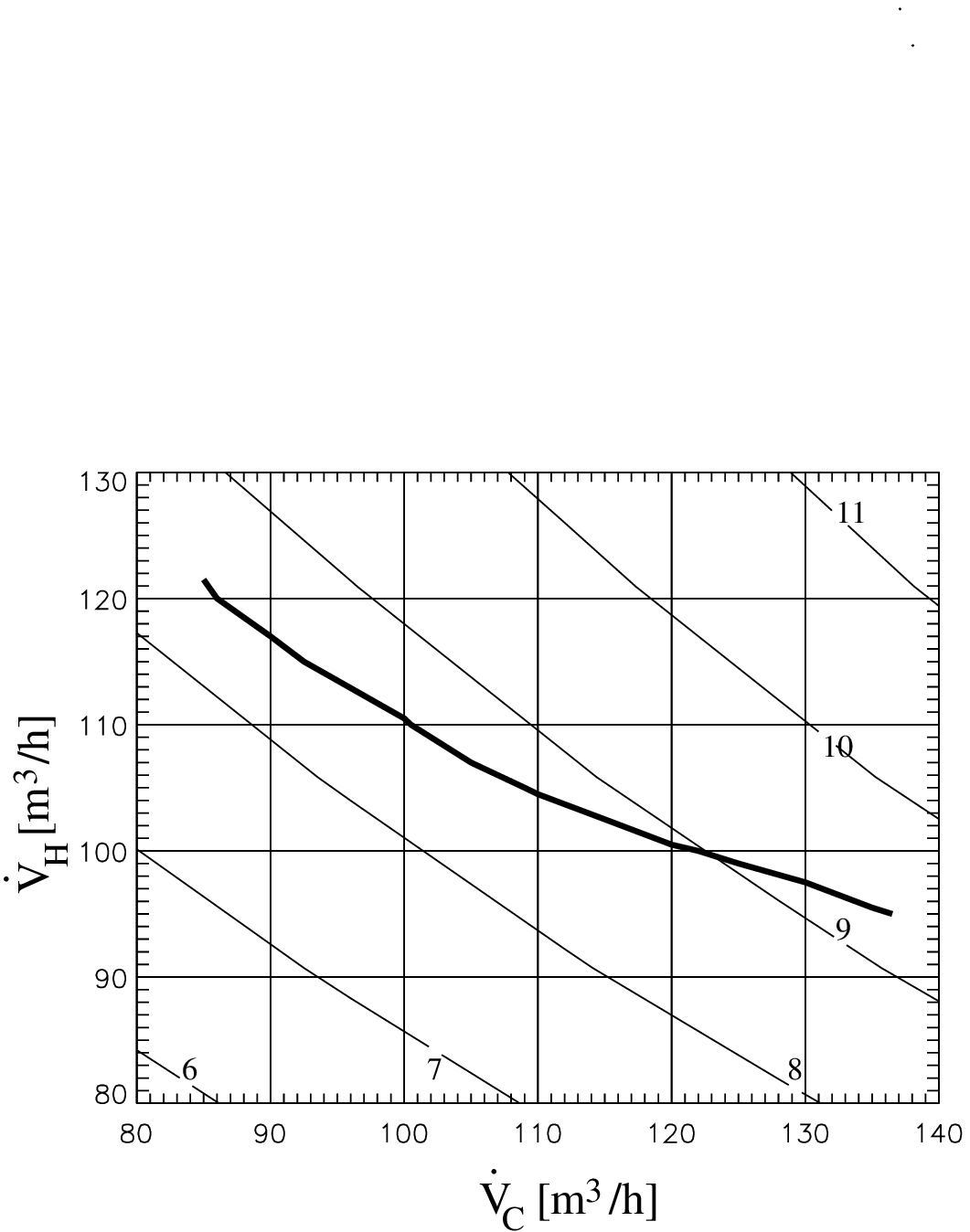}
\caption{Isolines of the dimensionless 
number $C=\mu_0 \sigma \alpha_{\perp} R$
in the $\dot{V}_C-\dot{V}_H$-plane. In a certain                        
approximation,                                                 
dynamo action should occur beyond the isoline with            
$C_{crit}=8.12$ \cite{RAEDLERMAHYD1}.
The                                                           
experimentally determined neutral                             
line (bold), separating regions with and without                
dynamo action,                                                   
slightly deviates from the theoretical line.                      
Figure courtesy of K.-H. R\"adler.}                              
\label{fig13}
\end{center}
\end{minipage}
\end{figure}

During its comparably short lifetime, the Karlsruhe dynamo
experiment has brought about
many results on its imperfect bifurcation behaviour and on
MHD turbulence which are documented in
\cite{MUST2,MUST3,MUST4}.                                                   
Much work has also been done in order to predict 
the kinematic dynamo behaviour  \cite{TILGNER97A,TILGNER97B,RAEDLERMAHYD1,TILGNER2002}
and  to                                      
understand quantitatively the saturated regime 
\cite{TILGNERBUSSE2001,RAEDLERMAHYD2,RAEDLERNPG}.
It should be mentioned that data from the Karlsruhe experiment have 
been used in an attempt to distinguish between two different scaling laws of the
geodynamo, 
with interesting consequences for its power consumption \cite{CHRISTENSENNATURE}.

A last remark: There is growing evidence that the disassembling of the 
Karlsruhe dynamo facility was too rash. Recent numerical simulations
have shown \cite{AVALOSALPHA} that just a very slight decrease of the 
aspect ratio of the dynamo module would have changed the
dominant dipole direction from equatorial to axial. It suggests 
itself  that in the vicinity of 
this point one could have expected quite interesting flip-flop phenomena
between equatorial and axial dipoles in the non-linear regime of the
dynamo. A similar point concerns transitions between 
steady and oscillatory dynamo regimes 
which typically occur at so-called 
exceptional points of the spectrum of the non-selfadjoint dynamo 
operator. Those transitions have been made responsible for the 
polarity reversals of the Earth's magnetic field \cite{REVERSALPRL,EPSL,GAFD}. 
In technical terms, an appropriate sign change of $\alpha$ along the radius
of the module would have been sufficient for such a transition to occur.
In any case, with modifying slightly the central module of the 
Karlsruhe dynamo, keeping all other parts of the
installation unchanged, there would have been a good chance to 
investigate very interesting effects.
Unfortunately, this opportunity has been missed.

\subsection{The VKS experiment in Cadarache}

At the CEA research center in Cadarache (France),
a group lead by J.-F. Pinton (ENS Lyon), S. Fauve (ENS Paris) and
F. Daviaud (CEA Saclay) has built a  dynamo experiment
under the acronym VKS (''von K\'arm\'an 
sodium''). Here, ''von K\'arm\'an''                                      
stands for the flow between two rotating disks \cite{ZANDBERGEN}.
\begin{figure}[h]
\begin{minipage}{75mm}
\begin{center}
\includegraphics[width=0.9\linewidth]{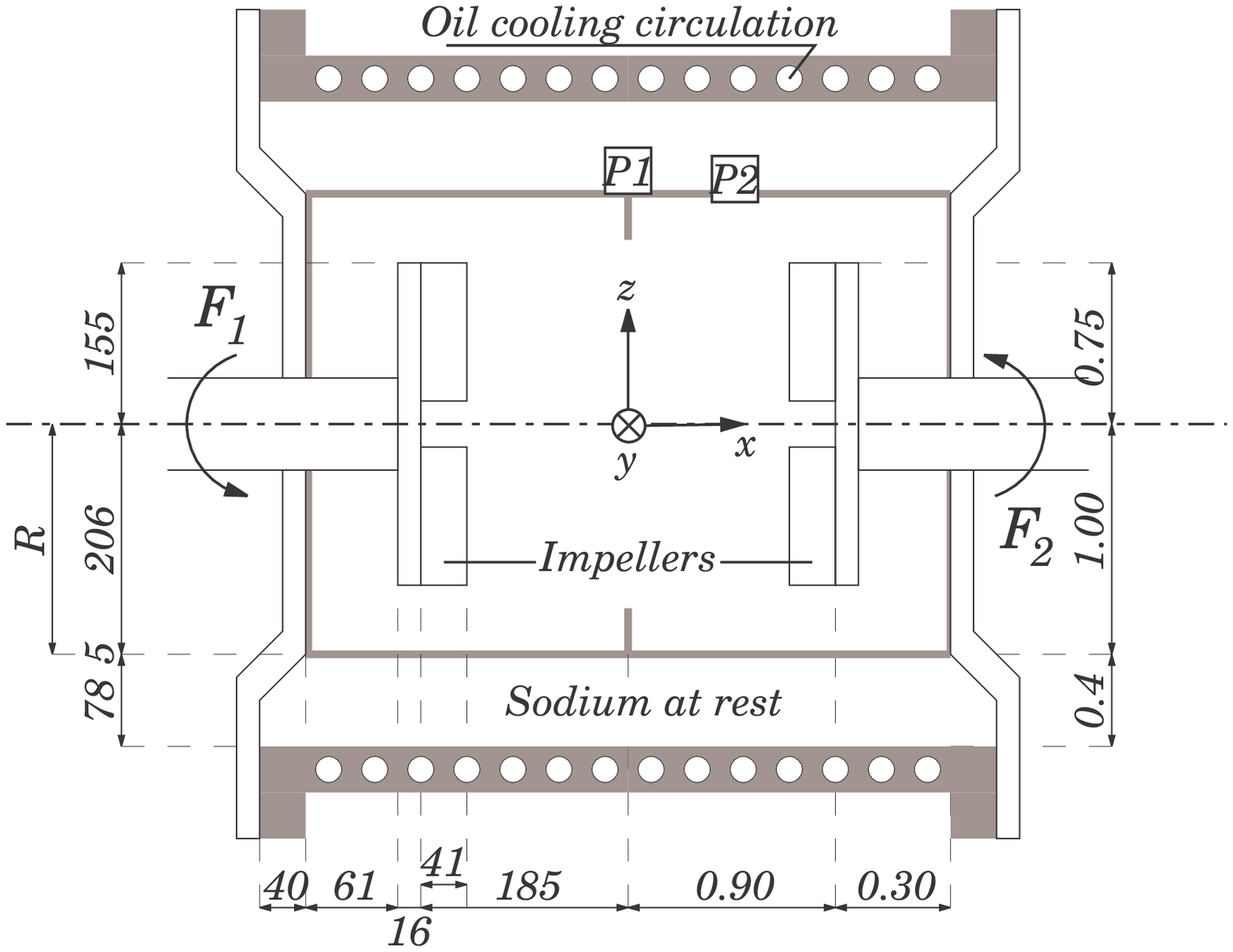}\\
\caption{Design of the VKS experiment 
with two disks counter-rotating in the cylinder driving two 
poloidal and two toroidal eddies (s2$^+$t2). The numbers on the l.h.s are 
the dimension in mm, the numbers on the
r.h.s are the dimensions normalized by the radius. 
Figure courtesy of the VKS team.}
\end{center}
\end{minipage}
\hfill
\begin{minipage}{75mm}
\begin{center}
\includegraphics[width=0.9\linewidth]{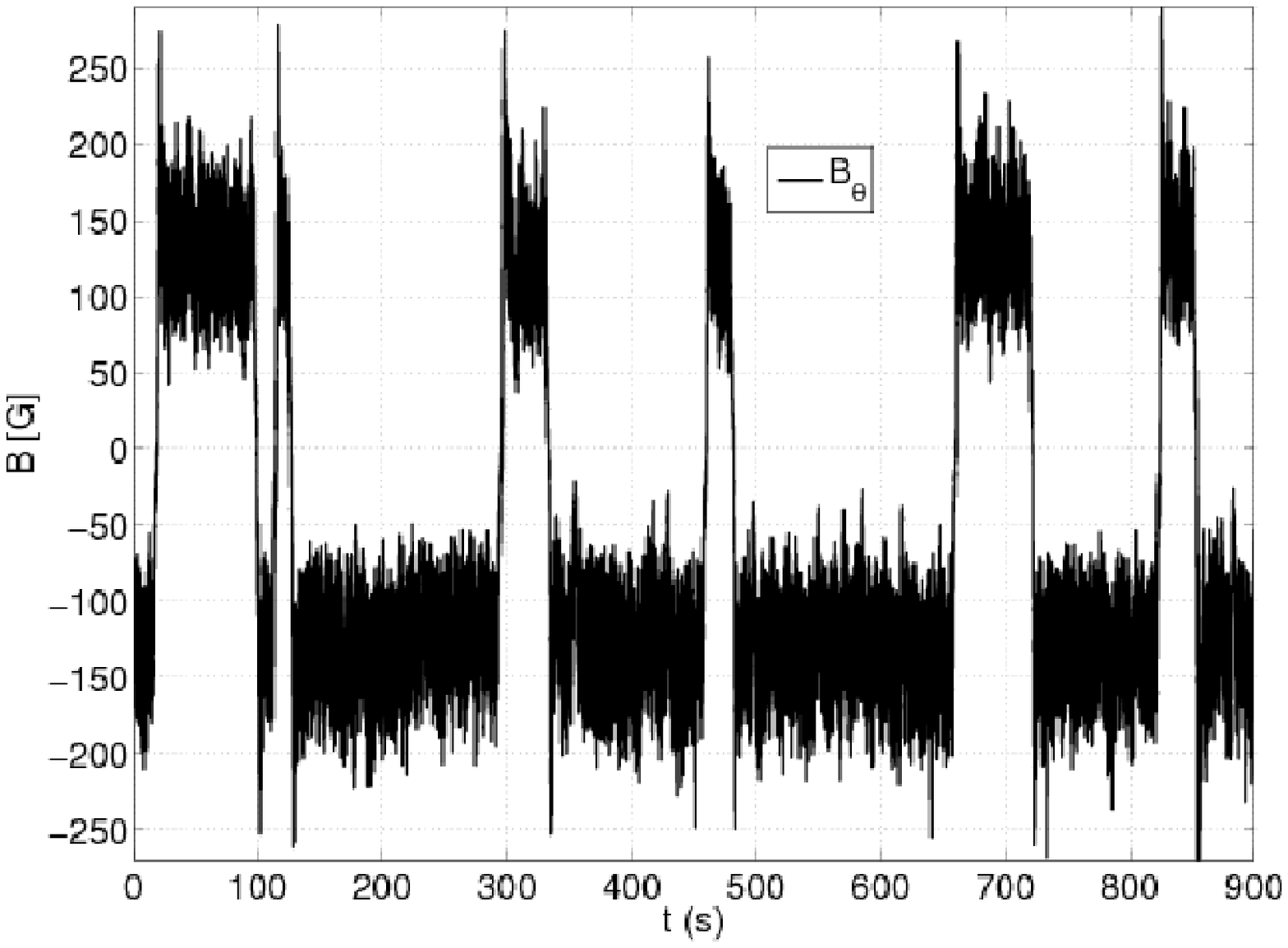}\\
\caption{A reversal of the azimuthal magnetic field occurring
when the rotation rate of one propeller was 16 Hz and that of the other 
propeller was  22 Hz.
Figure courtesy of the VKS team.}
\label{fig17}
\end{center}
\end{minipage}
\end{figure}

In  a first version of the experiment (VKS 1), the  flow was
produced inside a cylindrical vessel with equal diameter
and height, $2 R=H=0.4$ m,
driven by two 75 kW motors at rotation rates up to 1500 rpm. 
The von K\'arm\'an flow
geometry has been chosen as it represents a convenient 
realization of the so-called s2$^+$t2
flow that consists of two poloidal eddies (s2)
which are inward directed in the equatorial plane ($^+$), 
and two counter-rotating toroidal eddies (t2). 
Such a flow is known
to yield self-excitation at comparably low values of 
$Rm$ \cite{DUDLEYJAMES,NAKAJIMAKONO}. 
One problem of such flows is that the
counter-rotation in the
equatorial plane is a powerful source of turbulence.
Although significant induction effects were measured in this
first experiment 
\cite{MARIE2001,BOURGOIN,PETRELIS2002,PETRELIS2003},
no self-excitation was observed.

Based on this experience, a new version of the experiment (VKS 2) had been
constructed (Figure 12) . The total sodium volume was extended from 50 l to   %!!!
150 l, the available motor power from 75 kW to 300 kW, and                %!!!
great effort was spent
in order to optimize the shape of the blades of the impellers 
\cite{MARIE2003,RAVELET,MARIE2006}. 
Further, a side layer with sodium at rest was attached
which reduces the critical $Rm$ drastically. 
In spite of this thorough optimization, this experiment
failed to show self-excitation \cite{RAVELET2007}, and the induced
magnetic fields turned out to be significantly weaker than numerically
predicted \cite{RAVELETDISS}.
The reason for this general under-performance was controversially 
discussed. One ''school'' attributed it to 
(large scale) fluctuations \cite{LAVAL}, another one attributed it to 
the existence of ''lid layers'' behind the impellers, and 
in particular to the sodium rotation therein \cite{AMBI,JCP2007,LAGUERRE2006}.    %!!!!

In an attempt to mitigate this detrimental effect of lid layers, it was  
decided to change the axial magnetic boundary conditions by
replacing the stainless steel impellers by those made of soft-iron.
This modification led ultimately to the observation of
self-excitation in fall 2006 \cite{MONCHAUX}.
In some parameter regions, characterized by asymmetric forcing 
with different rotation rates 
of the two impellers, impressive magnetic field reversals 
(cf. Figure 13) were recorded \cite{BERHANU}.

The critical 
$Rm$ of the VKS 2 experiment with soft iron propellers 
was determined   to be $\approx 31...35$, in contrast
to the numerical predictions between 48 and 133 
(the latter depending mainly on the flow details in the lid layers).
The really surprising thing was, however, that
the self-excited eigenfield turned out to be basically an axisymmetric one,  %!!!
in contradiction to the non-axisymmetric equatorial dipole            %!!!
that was predicted by numerics.                                    %!!!
This axisymmetric field is apparently at odds with Cowling's    %!!!    
theorem \cite{COWLING} which forbids non-decaying axisymmetric    %!!!
eigenfields to be excited by large-scale flows.                   %!!!  

Hence, numerical work is going on to understand better the 
functioning of the VKS dynamo \cite{GISSINGER,GIESECKE2008}, 
in particular the role of the iron propellers
and the possible influence of helicity in the propeller region 
\cite{LAGUERRE2008}. A final explanation of the axisymmetric 
eigenmode  and its comparably low critical 
$Rm$ is, however, still missing.

\subsection{The Bullard-von K\'arm\'an experiment in Lyon}

A simple mechanical dynamo model, that had been proposed 
by Bullard in 1955, is the homopolar disk dynamo \cite{BULLARD55}:
Imagine a metallic disk rotating with an angular velocity $\bf{\omega}$
in a magnetic field $\bf{B}$. The emf $\bf{v} \times \bf{B}$
points from the axis to the rim of the disk and drives a current
$I$ through a wire that is wound around the axis of the disk. 
The orientation of the wire is such that
the external magnetic field is amplified. At a critical
value of $\bf{\omega}$, the amplification becomes infinite:
self-excitation sets in. With growing magnetic field, the
Lorentz force ${\bf{j}} \times \bf{B}$ acts against the driving
torque, which will ultimately lead to saturation. 
The feasibility of such an experiment was 
recently discussed in a paper by R\"adler and Rheinhardt \cite{RAEDLERDISK}.

Inspired by this disk dynamo, a team 
in Lyon has investigated  an interesting
experimental arrangement which generates
a magnetic field by a simple trick \cite{BOURGOINNJP}. 
Basically, it comprises a flow of the same s2$^+$t2 type
as the VKS flow. The working fluid is, however, 
gallium instead of sodium (''von K\'arm\'an gallium'' experiment, \cite{VKG}).   %!!!
The magnetic Reynolds number reaches only values
of 5 which is clearly not sufficient for dynamo action to occur.
The trick is now that a part of the amplification process is 
taken over by an external electrical amplifier which takes as input the 
azimuthal field component measured by a Hall sensor 
and feeds a coil which produces 
an axial magnetic field. Roughly speaking, the $\Omega$ effect (i.e. the    %!!!
generation of a toroidal from a poloidal field by means of             %!!!
differential rotation) is  produced by the flow, while the  $\alpha$    %!!!
effect is mimicked by the external amplifier.
Very interesting results have been obtained with this machine, including 
intermittent 
reversals of the dipole field, as well as excursions \cite{BOURGOINNJP}.

\subsection{Madison}

At the University of Wisconsin, Madison, C. Forest and his colleagues     %!!!
have undertaken the ''Madison dynamo experiment'' (MDX) 
which is, in many respects, quite similar
to the VKS dynamo. The flow topology is of the same s2$^+$t2 type with
two counter-rotating toroidal eddies and two
poloidal eddies which are pointing inward in the equatorial plane.
Historically it is noteworthy that Winterberg had proposed exactly 
this propeller configuration as early as 1963 \cite{WINTERBERG}.

The difference to VKS is that MDX works not in a cylinder but 
in a 1 m diameter sphere (Figure 14) and that 
the flow is driven by two impellers with K\"ort nozzles 
instead of two disks (with blades) as in VKS.                             %!!!
A lot of effort had been spent in the
hydrodynamic and numerical
optimization of the precise geometry
of the s2$^+$t2 flow \cite{FOREST}. 

\begin{figure}
\begin{minipage}{75mm}
\begin{center}
\includegraphics[width=0.9\linewidth]{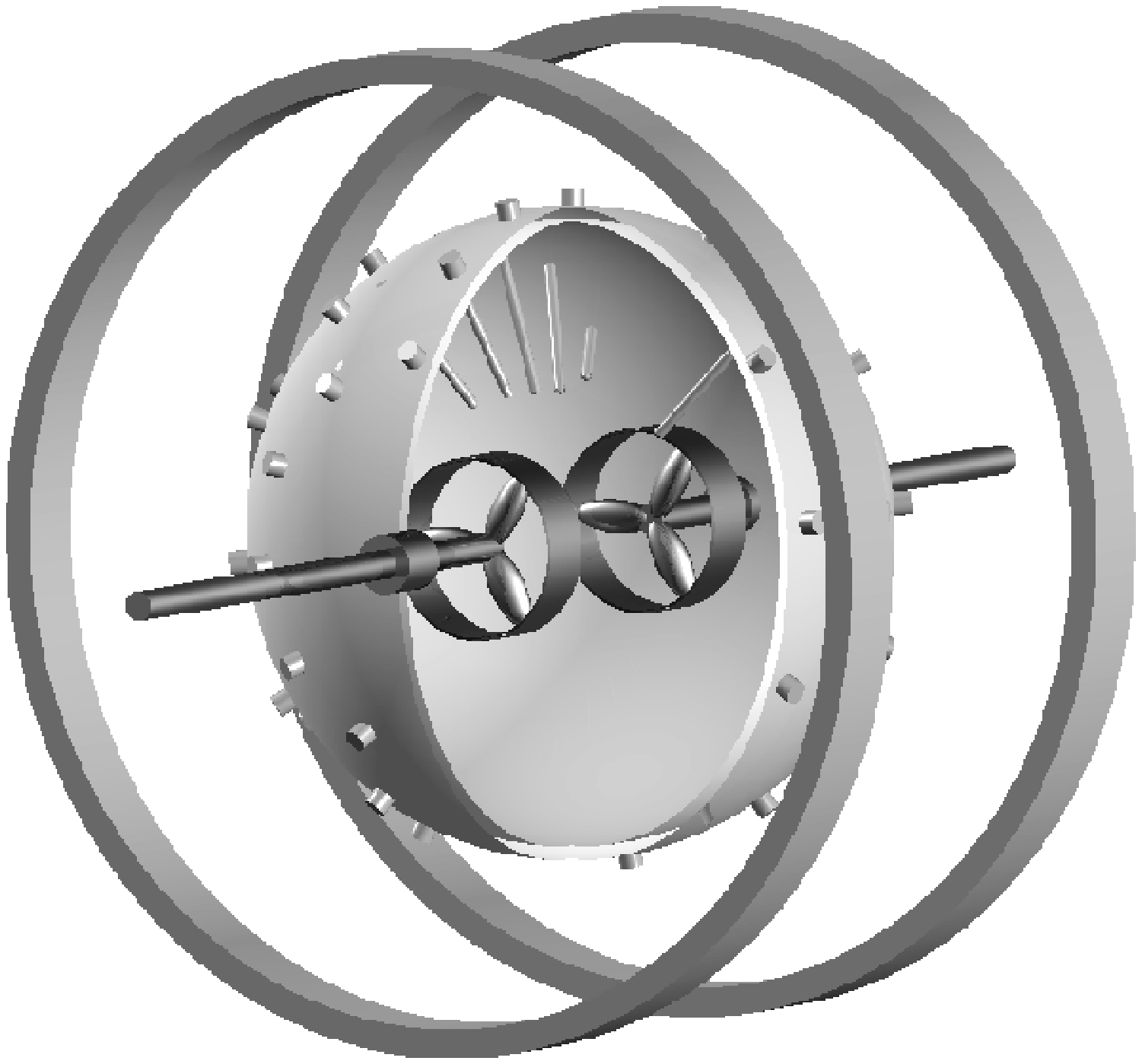}
\caption{Schematic plot of the Madison Dynamo experiment (MDX). The flow
topology is of the same s2$^+$t2 type as in the VKS experiment. 
The two rings symbolize  the Helmholtz coil for producing
an axial magnetic field. The lances for internal magnetic field
sensors are also shown. Figure courtesy of E. Spence.}
\end{center}
\end{minipage}
\hfill
\begin{minipage}{75mm}
\begin{center}
\includegraphics[width=0.9\linewidth]{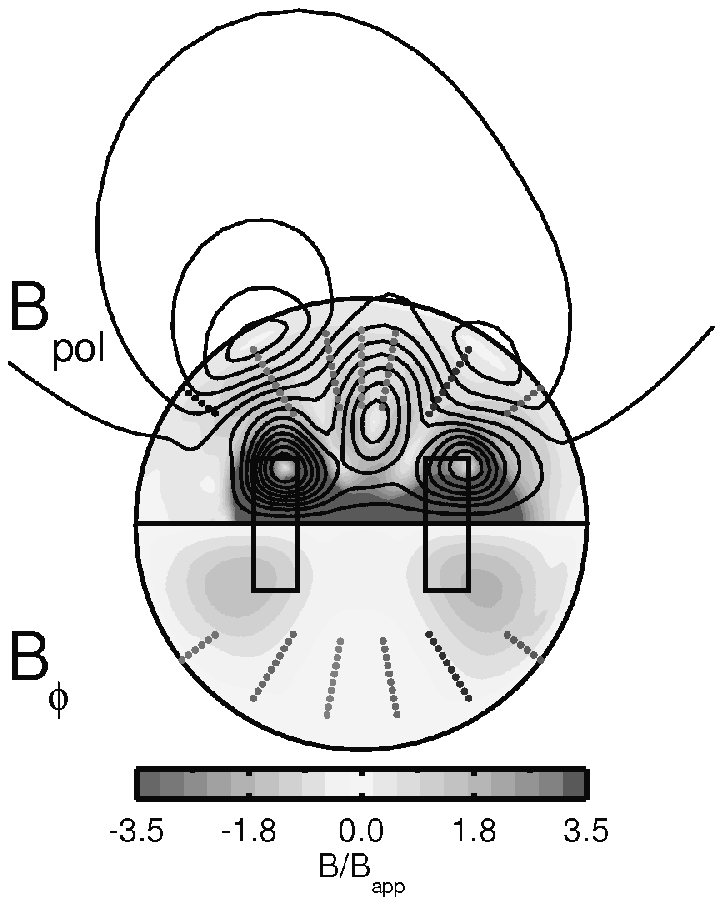}
\caption{Induced magnetic field structure measured in the MDX, including 
a significant axial dipole component.
Figure courtesy of E. Spence.}
\end{center}
\end{minipage}
\end{figure}

The MDX dynamo 
has not shown self-excitation up to present.
It was a surprise, however, that the measured {\it induced} magnetic field
turned out to be dominated by an axial dipole 
component (Figure 15), since
such an induced axial dipole cannot
be produced by a large scale axisymmetric flow \cite{SPENCEPRL1}.

Evidently, this  puzzle has a striking resemblance with the 
self-excitation of the axisymmetric field in the  VKS experiment.
The most plausible explanation comes from assuming some
sort of $\alpha$ effect in the flow. And again one has 
to bear in mind the extreme sensitivity of the mode selection
to minor amounts of helical turbulence (i.e. $\alpha$) 
in the impeller region that was
identified (for VKS) in \cite{LAGUERRE2008}. 

Further results have been published on 
intermittent 
bursts of dynamo action \cite{NORNBERGPRL}, the detailed measurement of
the magnetic field structure \cite{NORNBERGPLASMA}, 
and its interpretation in terms
of possible dynamo sources \cite{SPENCEPRL2}. In the latter paper,
the induction of an axial dipole has been interpreted as a sort of
''turbulent diamagnetism'', which is not necessarily in contradiction
with the interpretation given in \cite{LAGUERRE2008}.
Present activities at MDX point on further optimizing the flow
by installing various types of baffles.

Another focus of the Madison group is on replacing liquid metal
experiments by plasma experiments. A main step                          %!!!!
in this direction is to confine the plasma and to drive 
a rotating flow by means of crossed magnetic and electric fields
at the boundary \cite{BULLETIN1,BULLETIN2}. Interestingly,
this is a configuration which had been used                              %!!!!
in many flow control experiments with 
low conducting fluids \cite{GAI61,WEIER}.
The big advantage of plasma experiments is, of course, that 
$Pm$  is not a constant of the material but 
can be adjusted in a wide range \cite{WANGPARIEV}. 
By controlling the poloidal 
profile of the toroidal rotation, high $Rm$ flows will be generated 
that can result in MRI or  dynamo action.

\subsection{The rotating torus experiment in Perm}

An ingenious idea to circumvent the large driving power 
that is usually needed to do dynamo experiments has 
been pursued by the group of P. Frick 
at the Institute of Continuous
Media Mechanics in Perm, Russia 
\cite{FRICK1,PERMALPHA,FRICKPAMIR}. The idea 
relies on the fact
that a helical flow of the Ponomarenko type
can be
produced within a torus when its rotation is abruptly braked
and a fixed diverter forces the inertially continuing flow
on a helical path.
While this concept is very attractive not only with respect
to the low motor power that is necessary to
slowly accelerate the torus, but also with respect to the
fact that
the sodium can be perfectly confined in the
torus without any need
for complicated sealing, a less attractive feature is the
non-stationarity of the flow allowing only the study of a transient
growth and decay of a magnetic field.

\begin{figure}[h]
\begin{center}
\includegraphics[width=0.5\linewidth]{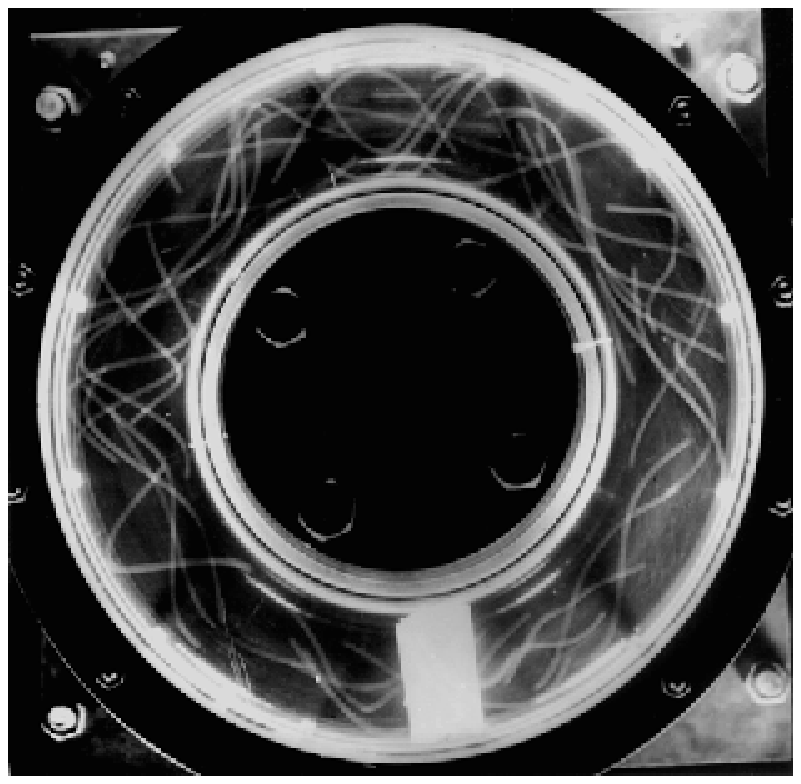}
\caption{Helical flow that develops after the abrupt brake
of the torus in the Perm water test experiment, visualized by               
polystyrene particles. The white bar at the bottom of the 
picture represents the diverter. Figure courtesy of P. Frick.}
\end{center}
\end{figure}

Extensive water pre-experiments and
numerical simulations \cite{DOBLERPERM} have been carried out to optimize and
predict magnetic self-excitation in such a non-stationary
dynamo. Figure (16) gives an impression of the flow that
appears shortly after the braking of the  torus.                  %!!!
The major radius of the water-filled torus is 10 cm, the              
minor radius is                                                       
2.7 cm. The photograph was taken 1.5 s after the full stop.

Based on these preparations a sodium experiment has been
build with the following
dimensions:
major radius of the torus 40 cm, minor radius of                       
the torus 12 cm,
mass of sodium 115 kg, rotation rate
3000 rpm, maximal velocity 140 m/s,
effective
magnetic Reynolds  number 40, minimal braking time
0.1 s. The main problem for such an experiments is connected with the 
tremendous mechanical stresses that appear in the short braking period.
A special bronze alloy has been used for of the torus.
First water experiments have been carried out, but 
for a sodium experiment more 
safety tests will be necessary.

In the preparatory phase of this experiment, 
important results on mean-field turbulence parameter
have been obtained with a smaller gallium 
experiment \cite{PERMALPHA,FRICKPAMIR}.
The importance of these parameters results from the question 
whether highly turbulent flows with high $Rm$                              %!!!
lead to an effective reduction of the
conductivity of the liquid. This reduction has been 
described in the context of mean-field dynamo theory as 
$\beta$ effect (cf. equation (11)).
While a significant reduction by a factor 10$^4$...10$^5$               %!!!
can be well justified for the solar dynamo \cite{KITCHATINOV2008},      %!!!
the corresponding value for the geodynamo is not safely known.             %!!!
Recent studies of reversal sequences and their statistical 
properties suggest that the effective conductivity of the Earth's outer core
might be reduced by a factor 3, when compared to the molecular 
conductivity of the material \cite{STOCHASTIC,IP}. 
Interestingly, this value is not that far from the 
factor 10 that was indicated by recent 
numerical simulations of mean-field coefficients in 
the geodynamo \cite{SCHRINNER2}.                        %!!!

While neither the Riga nor the Karlsruhe dynamo experiment  have shown
any measurable $\beta$ effect, there was only one experiment
in which the measurement of an 
$\beta$ effect had been claimed \cite{REIGHARDBROWN}.
Now, the Perm group has identified, by means of a very thorough measurement technique,
a conductivity reduction in the order of 1 per cent for a comparably low     %!!!!
$Rm$ of $\approx 1$ \cite{FRICKPAMIR}. This sounds not very much, 
but since  the dependence of $\beta$ on $Rm$ starts to be quadratic 
(in the low-conductivity limit)                                           %!!!
one could well
imagine a significant $\beta$ effect in planetary core flows characterized by 
much higher values of $Rm$.

\subsection{Sodium experiments in Maryland}

A variety of liquid sodium experiments have been carried
out under the guidance of D. Lathrop 
at the University 
of Maryland 
\cite{PEFFLEY1,PEFFLEY2,LATHROP2001,SHEW2001,SISAN2003,SISAN2004,SHEW2005}.  %!!

\begin{figure}
\begin{minipage}{75mm}
\begin{center}
\includegraphics[width=0.9\linewidth]{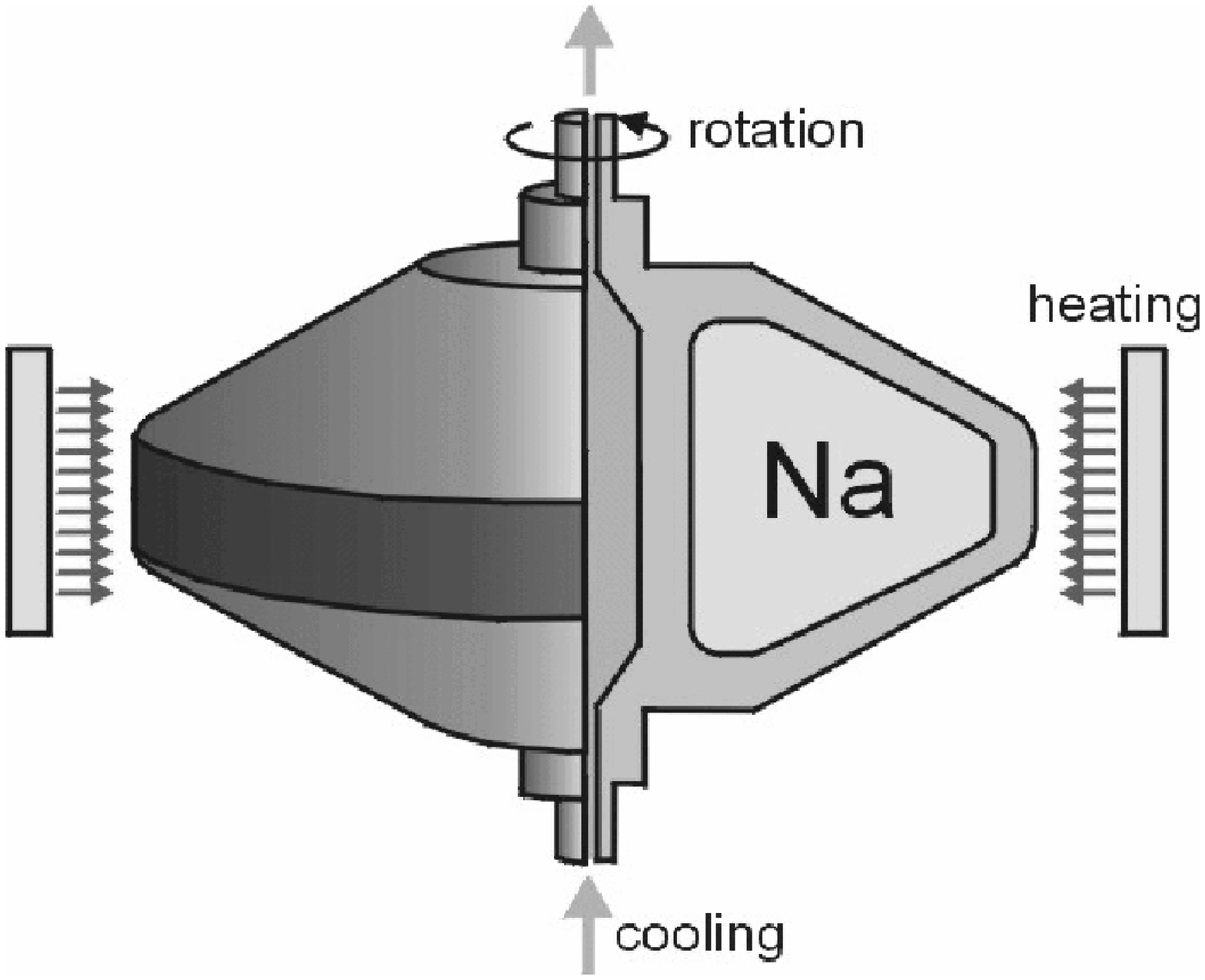}
\caption{The first dynamo experiment in Maryland. A rapidly rotating torus
is heated at the rim and cooled at the axis. Figure courtesy of D. Lathrop.}
\end{center}
\end{minipage}
\hfill
\begin{minipage}{75mm}
\begin{center}
\includegraphics[width=0.9\linewidth]{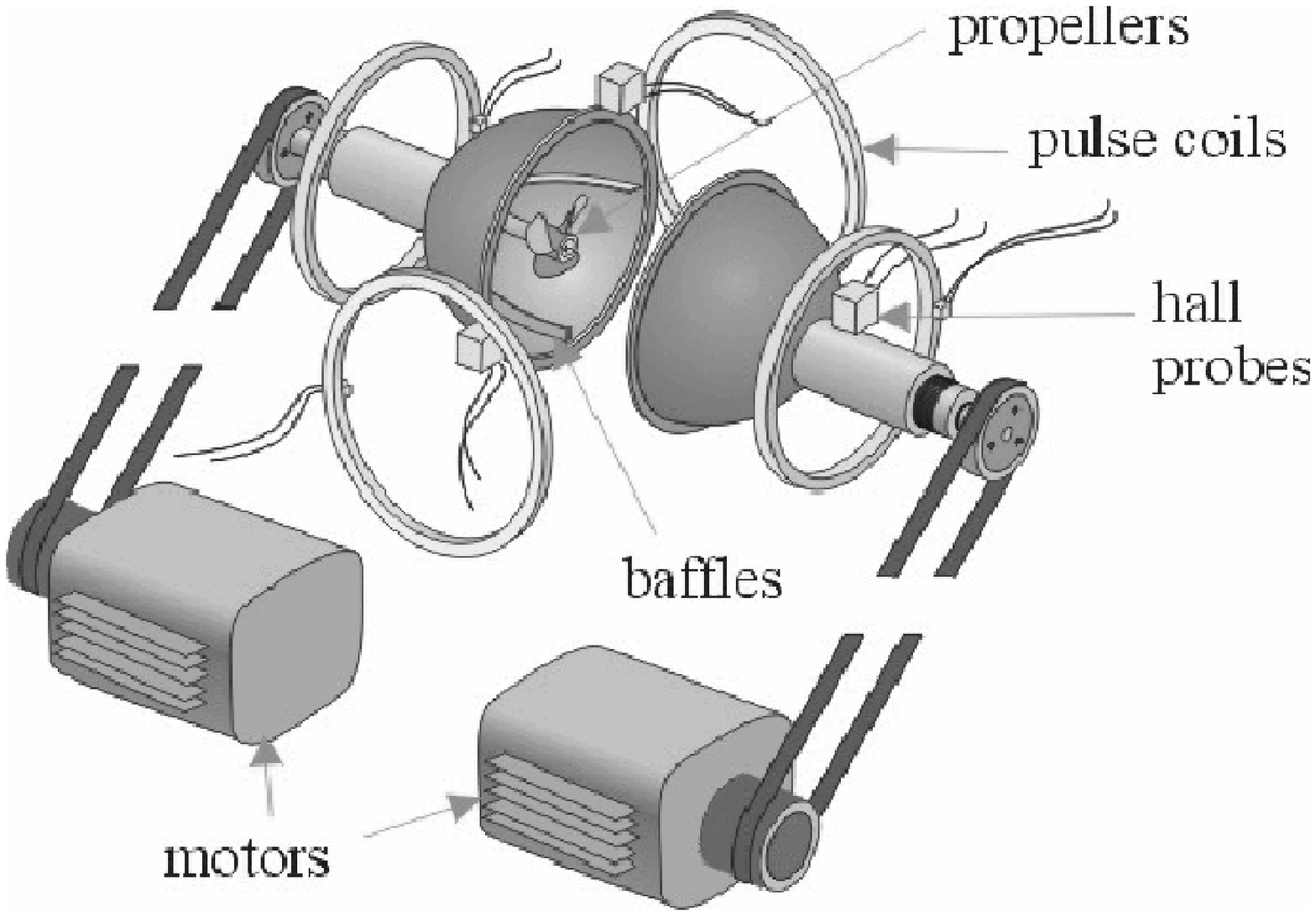}
\caption{The second dynamo experiment in Maryland. 
In a 0.3 m diameter sphere
different flows have been produced by propellers. 
This configuration, but with the propellers replaced by an
inner sphere, was used for the MRI experiment.
Figure courtesy of D. Lathrop.}
\end{center}
\end{minipage}
\end{figure}

In the first experiment (Dynamo I, see Figure 17),                        %!!!
a 0.2 m diameter titanium vessel containing 1.5 l of liquid sodium
was heated on the outer side and cooled at the axis.
The fast rotation (up to 25000 rpm!) was intended to induce
centrifugally driven convection, with the centrifugal 
force as a substitute for gravitation in the planetary case.
Self-excitation was not observed.

The second device (Dynamo II, see Figure 18)
consisted of a 0.3 m diameter sphere made of steel.
A total of  15 l of sodium was stirred by two counter-rotating 
propellers, each powered by 7.4 kW motors.
Note that the flow is again of the s2$^+$t2 topology as in VKS and MDX.
The most interesting result of this experiment was obtained by
carefully analyzing the decay rates of different modes.
The decay of an axisymmetric applied field
turned out to be more slowly as $Rm$ was increased, while
the decay of a non-axisymmetric field
even accelerated with increasing $Rm$. 
Actually the decrease of the decay rate of the axisymmetric 
magnetic field was by about
30 per cent compared to the un-stirred fluid,  at a magnetic
Reynolds number of about 65. Naively, a simple 
extrapolation of this trend would point to a critical $Rm$ of around 200.
Although we know from Cowlings theorem how problematic an interpolation
of decay rates of an axisymmetric mode towards zero is \cite{TILGNERMARYLAND}, 
one might ask again if not a similar mechanism as in VKS (and MDX) could
provide a  positive growth rate of an axisymmetric field.

\begin{figure}
\begin{minipage}{75mm}
\begin{center}
\includegraphics[width=0.9\linewidth]{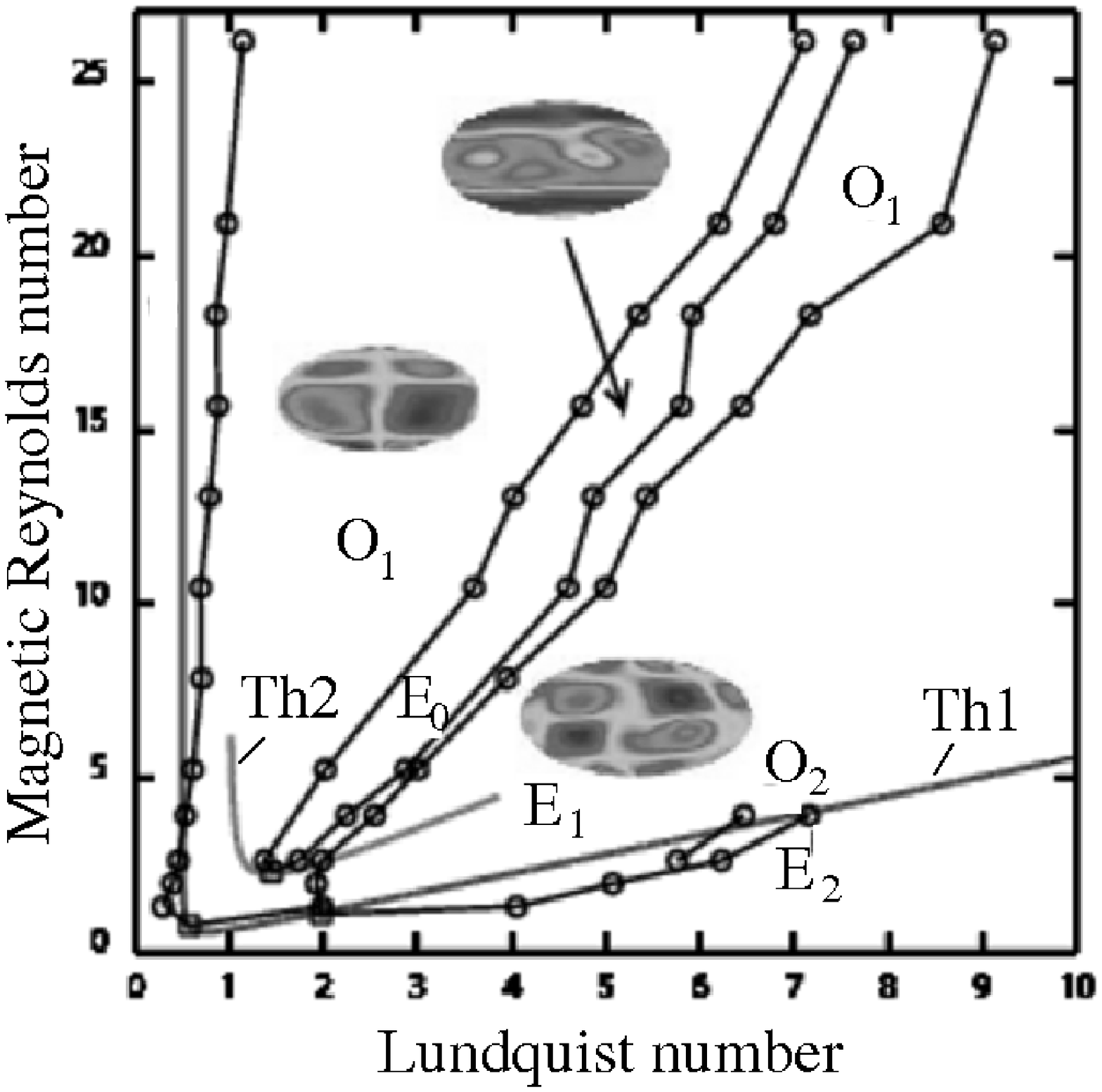}
\caption{Phase diagram  of the spherical Couette experiment in dependence on the 
Lundquist number and the magnetic Reynolds number. O1 E1 O2 AND e2 are different modes.
Th1 and Th2 are the theoretical stability boundary curves from the 
dispersion relation for the longest (Th1) and second longest
wavelength. Figure courtesy of D. Lathrop.}
\end{center}
\end{minipage}
\hfill
\begin{minipage}{75mm}
\begin{center}
\includegraphics[width=0.9\linewidth]{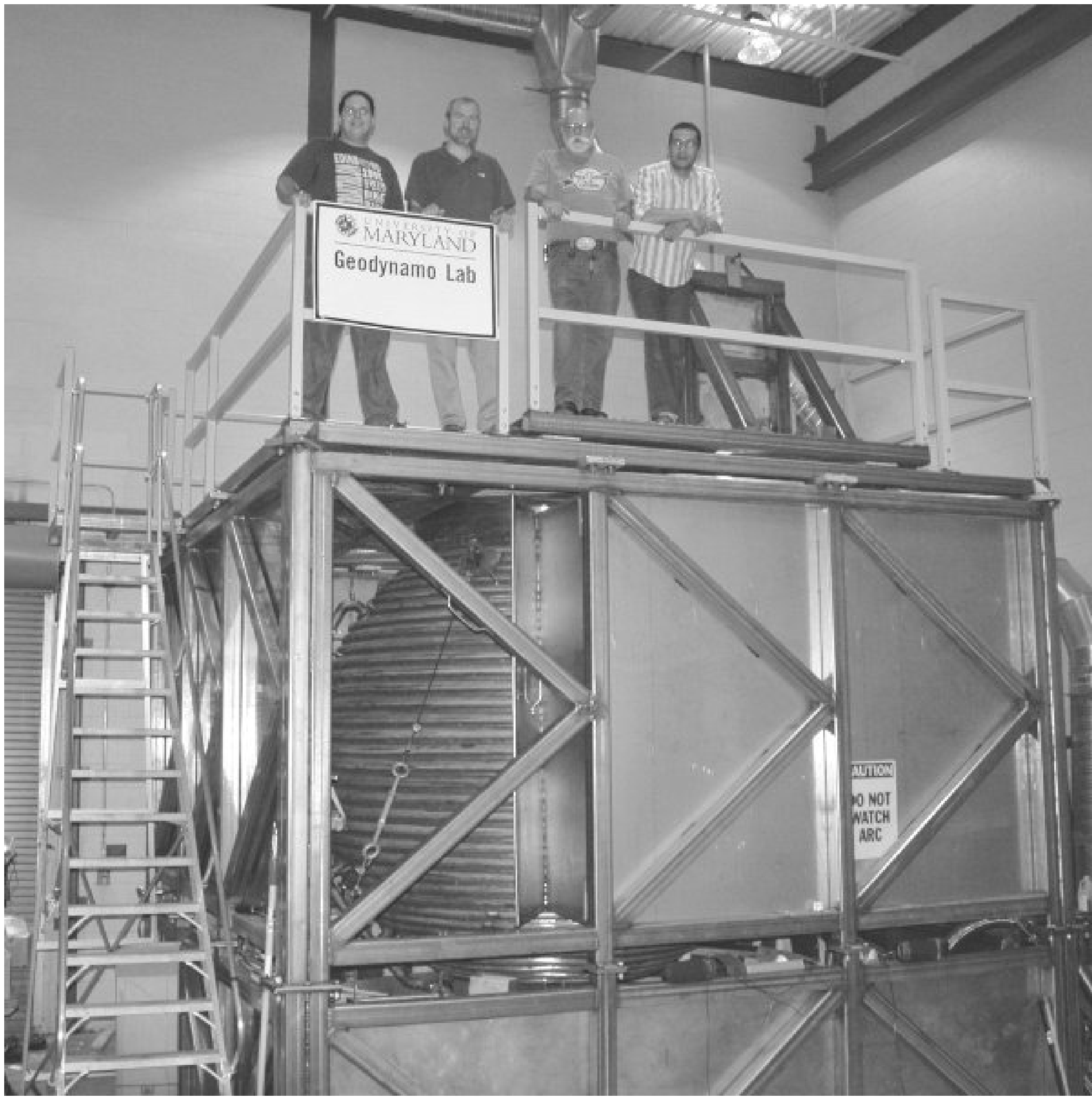}
\caption{The rotating 3m sphere in Maryland. An impressive video is 
available under www.youtube.com.
Figure courtesy of D. Lathrop.}
\end{center}
\end{minipage}
\end{figure}

Maybe the most important result of the Maryland group was obtained with a
modification of this Dynamo II. By replacing the two propellers by 
one 5 cm diameter sphere, one arrives at a classical spherical Couette
configuration. Applying, as before, an axial magnetic field to this
flow, new modes of correlated magnetic field and velocity perturbation 
appeared  \cite{SISAN2004} in certain parameter regions of $Rm$ and $S$ (see Figure 19).
On closer inspection
the dependence of these modes on $Rm$ and $S$ turned out  to be
in amazingly accurate correspondence with
predictions of the MRI based on the dispersion relations.
Running in an already strongly turbulent regime, this experiment
was certainly not able to show MRI as the {\it first instability 
of a stable flow}. 
Nevertheless, it is tempting to speculate (in the spirit of \cite{LIU1988}) 
that the fine-grained
background turbulence just gives some renormalized viscosity, 
without terribly affecting 
the basic mechanism of MRI that seems robust enough to show up 
as an ''coherent structure'' as long 
as only $Rm$  and $S$ are in the right range. 

The most recent project of the Maryland group is the 
construction of a giant rotating sphere with 3 m diameter (Figure 20), 
for future studies of MHD instabilities and (hopefully) the dynamo effect in
rotating systems.

\subsection{Grenoble}

Continuing the tradition of former geophysically inspired      
experiments \cite{BRITO1995,AUBERT2001},                       
a group in Grenoble has prepared the
so-called DTS (''Derviche Tourneur Sodium'') experiment
which is quite similar to the MRI experiment in Maryland.                       %!!!
The distinguishing feature is a permanent magnet in the
inner sphere in order to study the magnetostrophic regime \cite{CARDIN2002}       %!!!!
even when self-excitation is expectedly not achieved (Figure 21).

One of the most important results of the DTS experiment was the
experimental observation of strong super-rotation of the 
liquid sodium in the equatorial region \cite{NATAF2006}, which had been 
predicted by Dormy et al.                                                       %!!!!
in 1998 \cite{DORMY1998}. This was made possible by
inferring velocity profiles from
electric potentials measurements at the rim of the spherical container.
While the observed latitudinal variation of the electric potentials in 
the experiments 
differs markedly from the predictions of a numerical model similar to that of
Dormy, recent
numerical results show a better agreement with the measurements \cite{HOLLERBACH2007}.

A further focus of the DTS experiment is the investigation of 
various wave phenomena in magnetized rotating flows \cite{DENYS}.

\subsection{Princeton}

A group at Princeton Plasma Physics Laboratory, headed by H. Ji, 
has a long tradition in doing plasma experiments with relevance to
astrophysical processes, in particular to magnetic 
reconnection \cite{JIRECONNECTION} and the $\alpha$ effect \cite{JIDYNAMO} 
(which is, in the laboratory fusion plasma community, sometimes denoted as       %!!!
"dynamo effect" \cite{BLACKMANJI}).                                              %!!!

At present an experiment is under preparation that is intended to show
MRI in a Taylor-Couette cell filled with liquid gallium, at Reynolds numbers    %!!!
of several millions  \cite{JIGOODMANKAGEYAMA}.
The basic question for such an experiment is, of course, under which conditions 
the underlying flow
can really be considered laminar. Although Rayleighs criterion predicts
linear stability for the ratio of rotation rates of outer to inner cylinder being 
larger than the squared 
ratio of inner to outer radius,
it has long been thought that the effect of boundaries together with
non-linear instabilities will make the flow ultimately turbulent.
Actually, in the linear stable regime 
turbulence had been reported by a  number of authors
\cite{WENDT,SCHULTZGRUNOW,DUBRULLE}.

After a long optimization process \cite{KAGEYAMAJIGOODMAN}, the Princeton group 
has finally succeeded to find such a configuration of differentially rotating end rings
that preserves the Taylor-Couette profile of the angular velocity.
The measured torques, fluctuation levels and Reynolds stresses
suggest that the flow is indeed laminar up to Reynolds numbers of about 2x10$^6$
\cite{JINATURE}. Experiments with liquid gallium are presently under preparation.
Recent numerical simulations
suggest that the MRI should indeed be identifiable in such an experiment,
in particular by its linear growth and the increased torque \cite{LIU2008}.

However, as usual for such experiments, some uncertainties remain.
They concern, in particular, the not well understood role of 
the rotating rings in the end-caps. Actually, the rotation rates which have been
chosen in the experiment in order to restore the Taylor-Couette flow profile 
are different from the numerically optimized ones. To explain this discrepancy, 
even cylinder 
wobbling has been invoked, which may point to a quite complicated process
involved in restoring the Taylor-Couette profile \cite{ROACH,LIU2008}.

On the other hand, the Maryland experiment has shown that MRI seems to be a quite
robust phenomenon that appears quite independently on other
flow features as long as only the necessary combination of $Rm$ and $S$ 
is reached.

However this might be, the Princeton gallium experiment will certainly teach us a lot
about hydromagnetic instabilities in Taylor-Couette flows at high Rm.

\begin{vchfigure}[h]
\begin{minipage}{75mm}
\begin{center}
\includegraphics[width=0.9\linewidth]{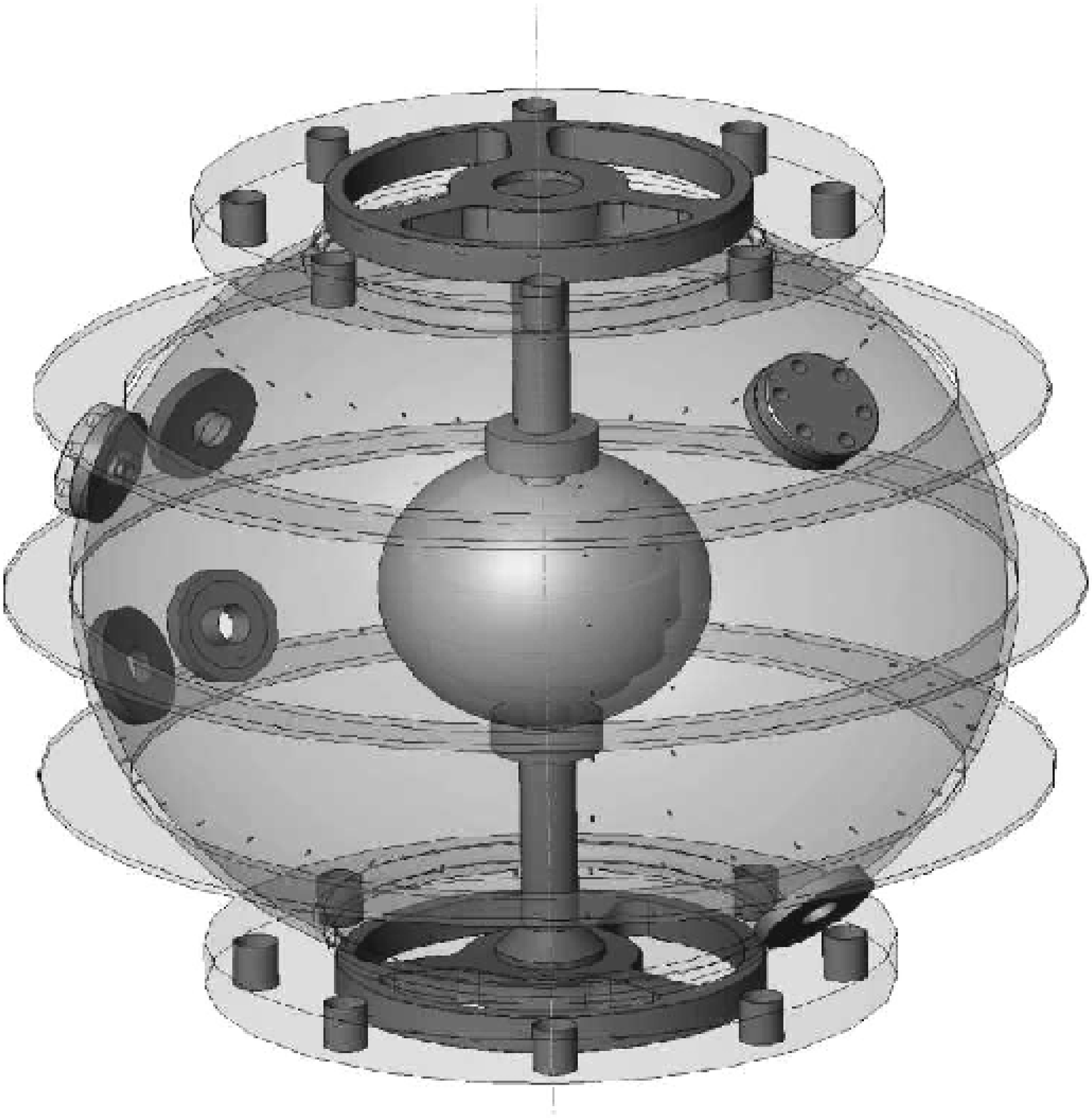}
\caption{The DTS experiment in Grenoble. Forty liters of liquid sodium are
contained between a 7.4 cm inner sphere and a 21 cm radius outer sphere.
The copper inner sphere contains a magnet which produces a nearly dipolar field
with a maximum of 0.345 T in fluid close to the poles. Figure courtesy of D. Schmitt and
the DTS team.}
\end{center}
\end{minipage}
\hfill
\begin{minipage}{75mm}
\includegraphics[width=0.9\linewidth]{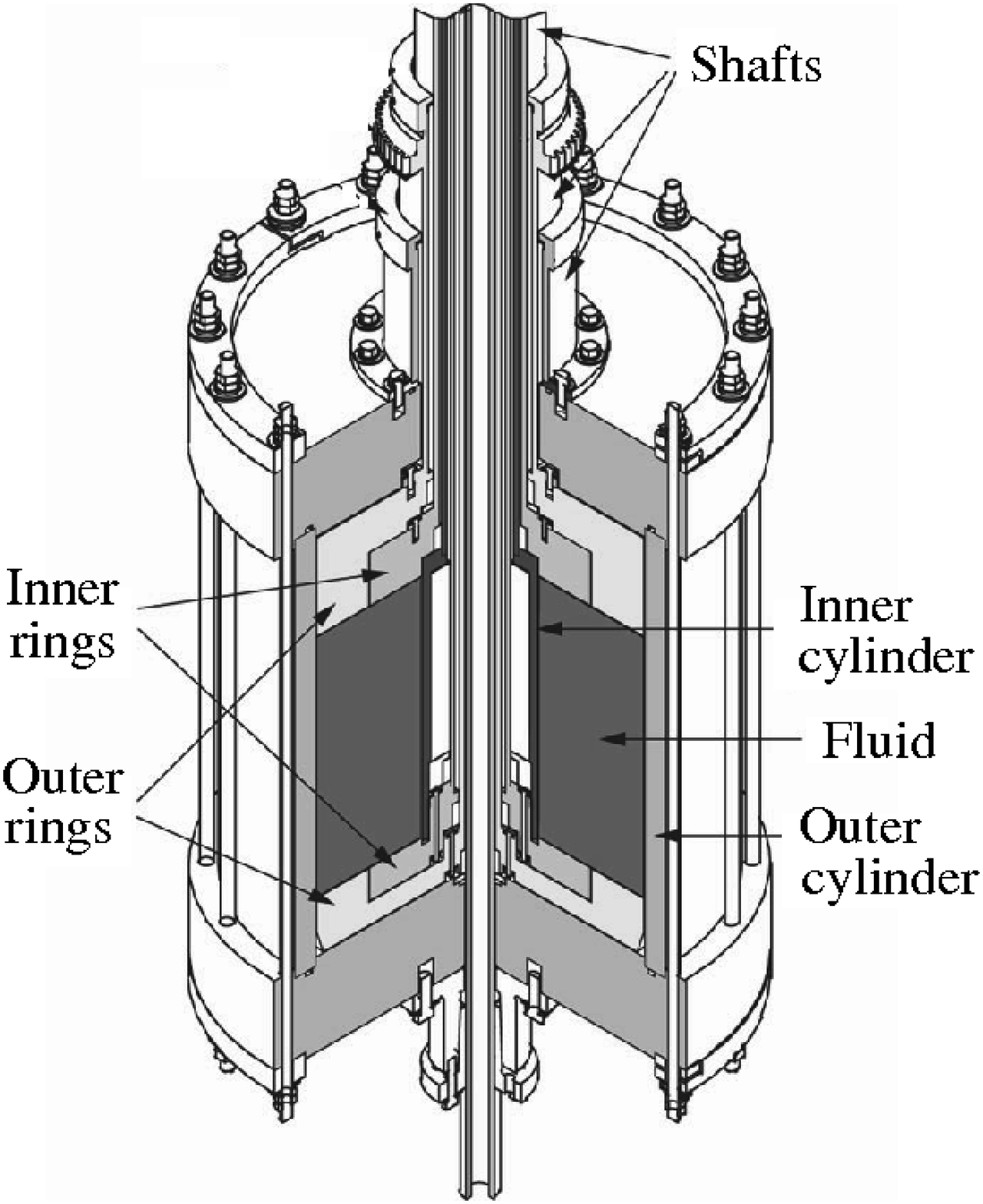}
\caption{The Princeton MRI experiment. 
The rotating gallium of height 27.86 cm is contained
between two differentially rotating 
concentric cylinders of radii 7.06 cm and 
20.30 cm. Figure courtesy of H. Ji.}
\end{minipage}
\end{vchfigure}

\subsection{The sodium experiment in New Mexico}

Since a couple of years, a sodium experiment 
is under construction at the New Mexico
Institute of Mining and Technology (NMIMT) in Soccoro          %!!!
\cite{COLGATE2001,COLGATE2002}.
The title of the
project is ``The $\alpha- \Omega$ accretion disk dynamo
that
powers active galactic nuclei (AGN) and creates the 
magnetic field of the universe.''

At the present stage, the experiment is a 
Taylor-Couette experiment (Figure 23),
quite similar to the Princeton experiment.                             %!!!
It is planned to produce an $\alpha$ effect by                         %!!!!
means of 
''plumes''   that are 
driven by pulsed jets.
The envisioned $Rm$, based on the
rotation alone, is 130, the corresponding  $Rm$ for the plumes
is 15 \cite{COLGATE2001}.
The water experiments have already                                   
revealed that 
the differential rotation of the Couette flow speeds up               
the anticyclonic rotation of the plumes.                              
This anticyclonic rotation will form the basis for the                    
$\alpha$-effect of the $\alpha$-$\Omega$-dynamo.

\begin{vchfigure}[h]
\begin{center}
\includegraphics[width=0.95\linewidth]{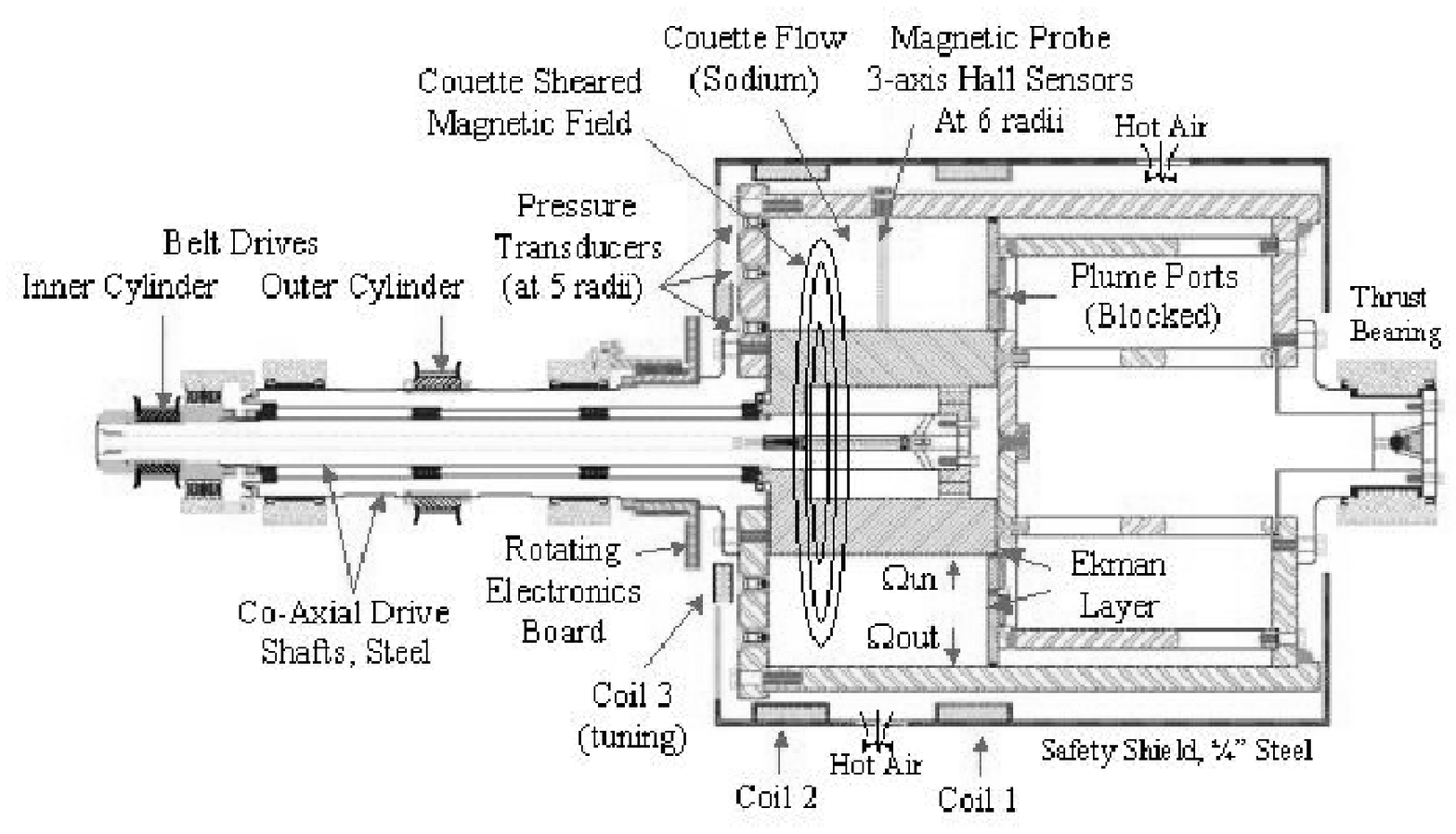}
\caption{The NMIMT $\alpha-\Omega$ experiment, at the present stage without   %!!!
the pulsed jet. Figure courtesy of S. Colgate.}
\end{center}
\end{vchfigure}

\subsection{The MRI experiment in Obninsk}

An MRI experiment with liquid sodium, proposed by 
E.P. Velikhov, is under preparation at the Institute of Physics and Power 
Engineering 
in Obninsk (Russia), in collaboration with the Kurchatov Institute in Moscow.
The basic idea is to drive a flow in a torus of rectangular
cross-section by a Lorentz force due to an applied vertical field and
an applied radial current \cite{VELIKHOV2006a,KHALZOV2006a,KHALZOV2006b}.
The hope is that in the bulk of the volume one gets
an angular velocity dependence $\sim r^{-2}$ which would exactly 
correspond to the Rayleigh line.
A similar experiment had been proposed in \cite{DEAN}, although as a Taylor-Dean flow
to have the angular momentum increasing at all radii.
While experimental results are not available from those experiments
an interesting claim has been made on a re-interpretation of the
Moresco-Alboussiere experiment \cite{MORESCO,KRASNOV} on the stability 
of the Hartmann flow in terms
of MRI \cite{KHALZOV2007}.

\subsection{The PROMISE experiment in Dresden-Rossendorf}

In this last subsection we leave the realm of high $Rm$ flows
and discuss an experiment on a particular type of MRI which has been coined 
''inductionless MRI'' \cite{PRIEDE1}
or ''helical MRI'' (HMRI) \cite{LIUJIGOODMAN}.

The background for this is the following:
We have seen, in connection with ''standard MRI'' 
(SMRI) experiments with  an axially applied magnetic field,
that it is extremely difficult to keep those high $Re$ 
flows laminar. However, those high $Re$
are not a genuine necessity for MRI,
but just a consequence  of the need for $Rm$
in the order of one or larger.

The reason for this is that the azimuthal magnetic field 
(which is an  unavoidable ingredient for MRI) must be 
{\it induced}  from the 
applied axial field by the rotation of the flow, and such 
induction effects are just proportional to $Rm$.
This being said, one could ask why not to replace
the induction process by just {\it externally applying} 
the azimuthal magnetic field as well? This question was addressed
in a 2005 paper by Hollerbach and R\"udiger \cite{HORU}, who showed 
that the MRI is then possible with dramatically reduced experimental
effort. Actually, the scaling characteristics of this HMRI 
are completely different from  those of SMRI.
While the latter needs  $Rm$ and $S$
of the order of 1, the former depends on $Re$ and
$Ha$ (or the interaction parameter $N$).

HMRI  is currently the subject of intense discussions in the 
literature 
\cite{PESSAHPSALTIS,RUEDIGERAN,LIUGOODMANHERRON,LIUJIGOODMAN,PRIEDE1,
ANJACEK1,ANJACEK2,LAKHIN2007},
the roots of which trace back to an early dispute between Knobloch
\cite{KNOBLOCH1} and Hawley and Balbus \cite{BAHA2}.

A remarkable property of HMRI, which has been clearly worked out in
\cite{PRIEDE1}, is the apparent paradox that a
magnetic field triggers an instability 
though the dissipation is larger than without magnetic fields. 
This is not so surprising 
when put in the context of other {\it dissipation induced instabilities} which 
are quite common
in many areas \cite{RMPDISSIPATION,KIRILLOV}.

Another, and not completely resolved, issue concerns the relevance
of HMRI for astrophysical flows. On first glance, HMRI seems well
capable to work in cold regions of accretion disks characterized by
small $Pm$ where SMRI cannot work. This scenario
might indeed be important for the ''dead zones'' 
of protoplanetary disks \cite{DEADZONES}
as well as for the outer parts of accretion disks around black holes
\cite{BALBUSPM}. 

However, before entering such a discussion in detail, one has to check whether HMRI
works at all for Keplerian rotation profiles $\Omega(r)\sim r^{-3/2}$. 
While the answer resulting from the dispersion relation was negative 
\cite{LIUGOODMANHERRON}, the solution of the eigenvalue equation gave an
affirmative answer,
as long as at least the outer or the inner radial boundary is conducting
\cite{HOLLERBACHRUEDUGERPRE}.

Unfortunately, even this is not the end of the story. Since HMRI appears 
in the form of a travelling wave, one has to be quite careful with 
the interpretation of the instability of a single monochromatic wave. 
Actually, 
one has to look for wave packet solutions with vanishing
group velocity. Typically, the regions in parameter
space for this {\it absolute instability} are only a subset of those for the
{\it convective instability}. A comprehensive analysis of this topic
is under preparation \cite{PRIEDE2}.

Notwithstanding this ongoing discussion, the dramatic decrease of the
critical $Re$ and $Ha$  for the onset of the MRI in helical magnetic
fields, as compared with the case of a purely axial field, made this new
type of MRI very attractive for experimental studies.

\begin{vchfigure}[h]
\begin{center}
\includegraphics[width=0.9\linewidth]{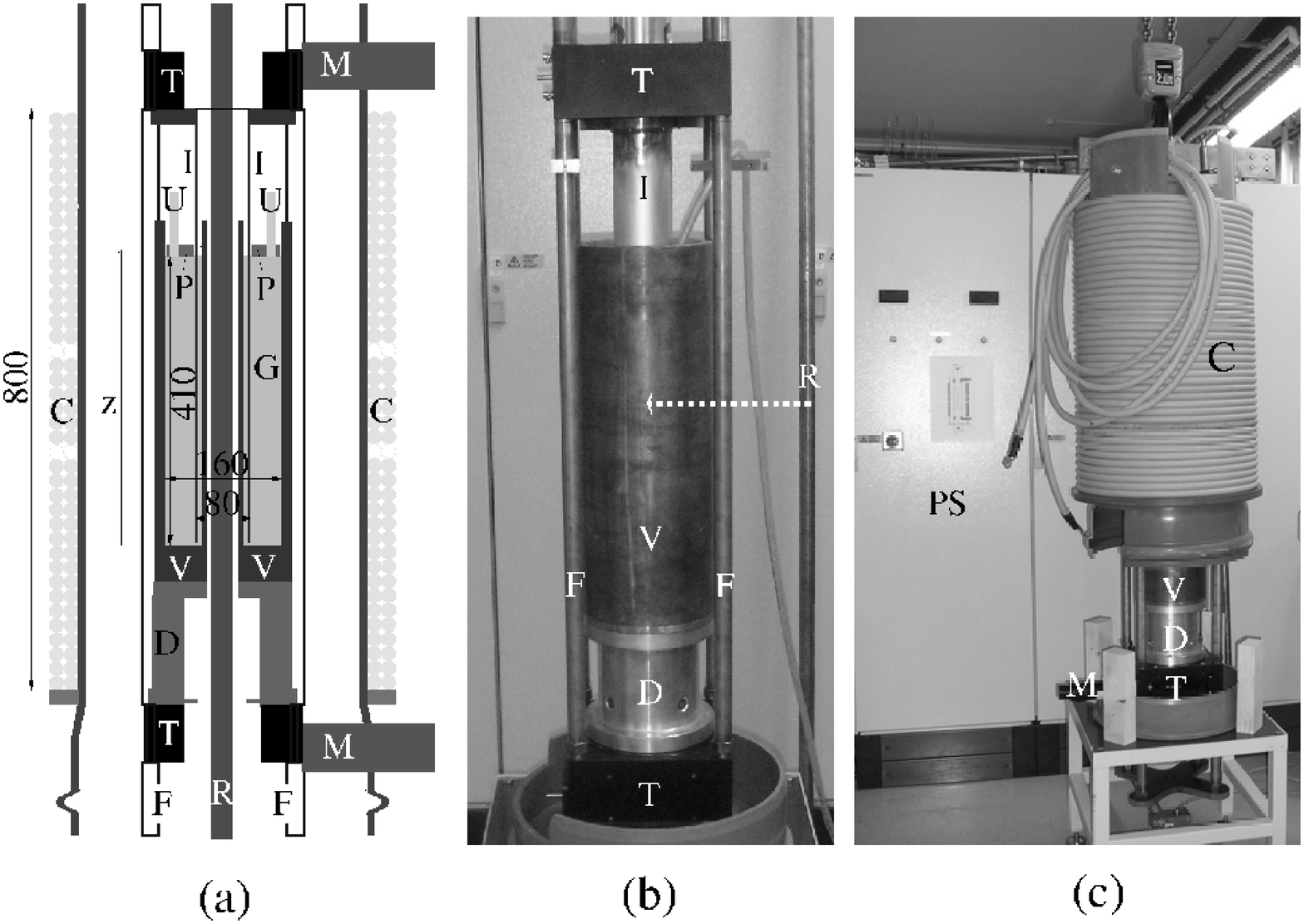}
\caption{The PROMISE experiment. (a) Sketch. (b) - Photograph of the 
central part. (c) - Total view with the coil being installed. V - Copper vessel, 
I - Inner cylinder, G - GaInSn, U -  Two ultrasonic transducers, 
P  - Plexiglas lid, T - High precision turntables, M - Motors, 
F - Frame, C - Coil, R - Copper rod, PS - Power supply for currents up to 8000 A.  
The indicated dimensions are in mm.}
\end{center}
\end{vchfigure}

The PROMISE facility ({\it P\/}otsdam {\it RO\/}ssendorf
{\it M\/}agnetic {\it I\/}n{\it S\/}tability {\it E\/}xperiment),
shown in Figure 24, is basically a cylindrical Taylor-Couette
cell with externally imposed axial and azimuthal magnetic
fields. Its primary component is a cylindrical copper vessel V, fixed on
a precision turntable T via an aluminum spacer D.  The inner wall of this
vessel is 10 mm thick, extending in radius from 22 to 32 mm; the outer
wall is 15 mm thick, extending from 80 to 95 mm.  The outer wall of this
vessel forms the outer cylinder of the TC cell.  The inner cylinder I, also
made of copper, is fixed on an upper turntable, and is then immersed into
the liquid metal from above. It is 4 mm thick, extending in radius from
36 to 40 mm, leaving a 4 mm gap between it and the inner wall of the
containment vessel V. The actual TC cell therefore extends in radius from
40 to 80 mm, for a gap width $d=r_o-r_i=40$ mm.  This amounts to a radius
ratio of $\eta:=r_i/r_o=0.5$. The fluid is filled to a   
height of 400 mm, for an aspect ratio of $\sim\!10$.
Axial fields of the order of 10 mT are produced by a coil C, and azimuthal 
fields of the same order are produced by currents of up to 8000 A in a 
water cooled copper rod R going through
the center of the facility.

First results of PROMISE
were published recently \cite{MRIPRL,MRIAPJL,MRINJP}.
The  most important result was the appearance of MRI in form of a travelling wave 
in a limited window of $Ha$ (which is proportional to the coil current $I_{coil}$).
The frequency of this wave turned out to be in good 
agreement with numerical predictions.

The intricacies of this PROMISE 1 experiment, as we call it now,
trace back to the two points 
discussed above. First, HMRI only slightly extends beyond the Rayleigh line 
and, second, the small unstable region for the convective instability
is further reduced when considering the absolute instability.
From this  high sensitivity of  the instability
with respect to $\mu:=\Omega_{o}/\Omega_{i}$  one has to be careful with Ekman/Hartmann pumping and
a possible change of the profiles by short circuited currents which can drive a 
Dean flow \cite{ANJACEK2}.
Another point is the critical  role of the radial jet approximately at mid-height
of the Taylor-Couette cell at which the MRI wave is typically  stopped \cite{STEFANIAN}. This
radial jet results from the Ekman pumping at the lower and upper end caps.

\begin{vchfigure}
\begin{minipage}{65mm}
\begin{center}
\includegraphics[width=0.95\linewidth]{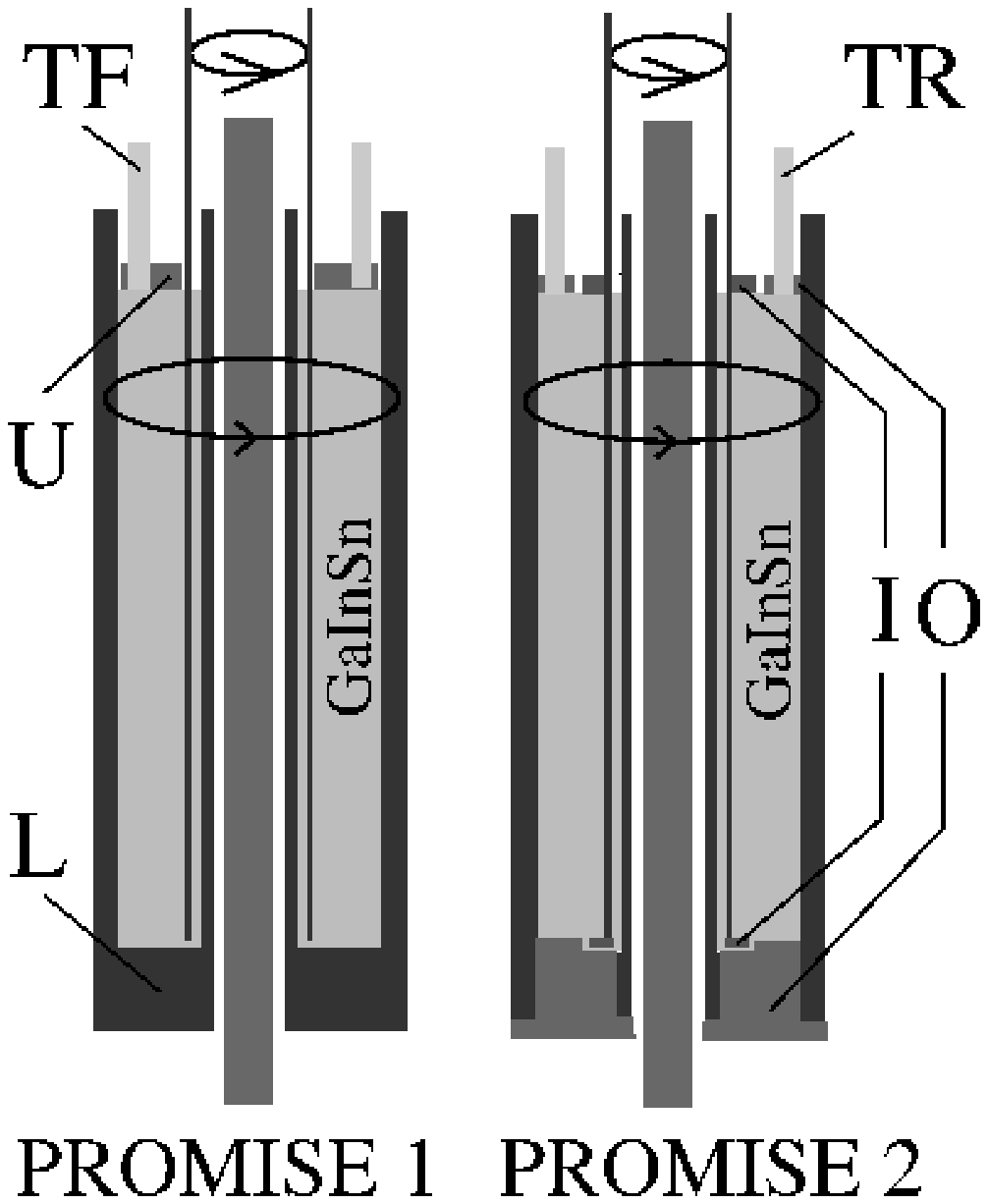}
\caption{The asymmetric end cap configuration in 
PROMISE 1 is replaced by a symmetric one in PROMISE 2. U - Upper 
end cap fixed in the laboratory frame, L - Lower end cap, made of copper, rotating with the 
outer cylinder, I -  Inner plastic rings, O - Outer plastic rings, 
TF - Fixed ultrasonic transducer, TR - Rotating ultrasonic transducer.}
\end{center}
\end{minipage}
\hfill
\begin{minipage}{82mm}
\begin{center}
\includegraphics[width=0.95\linewidth]{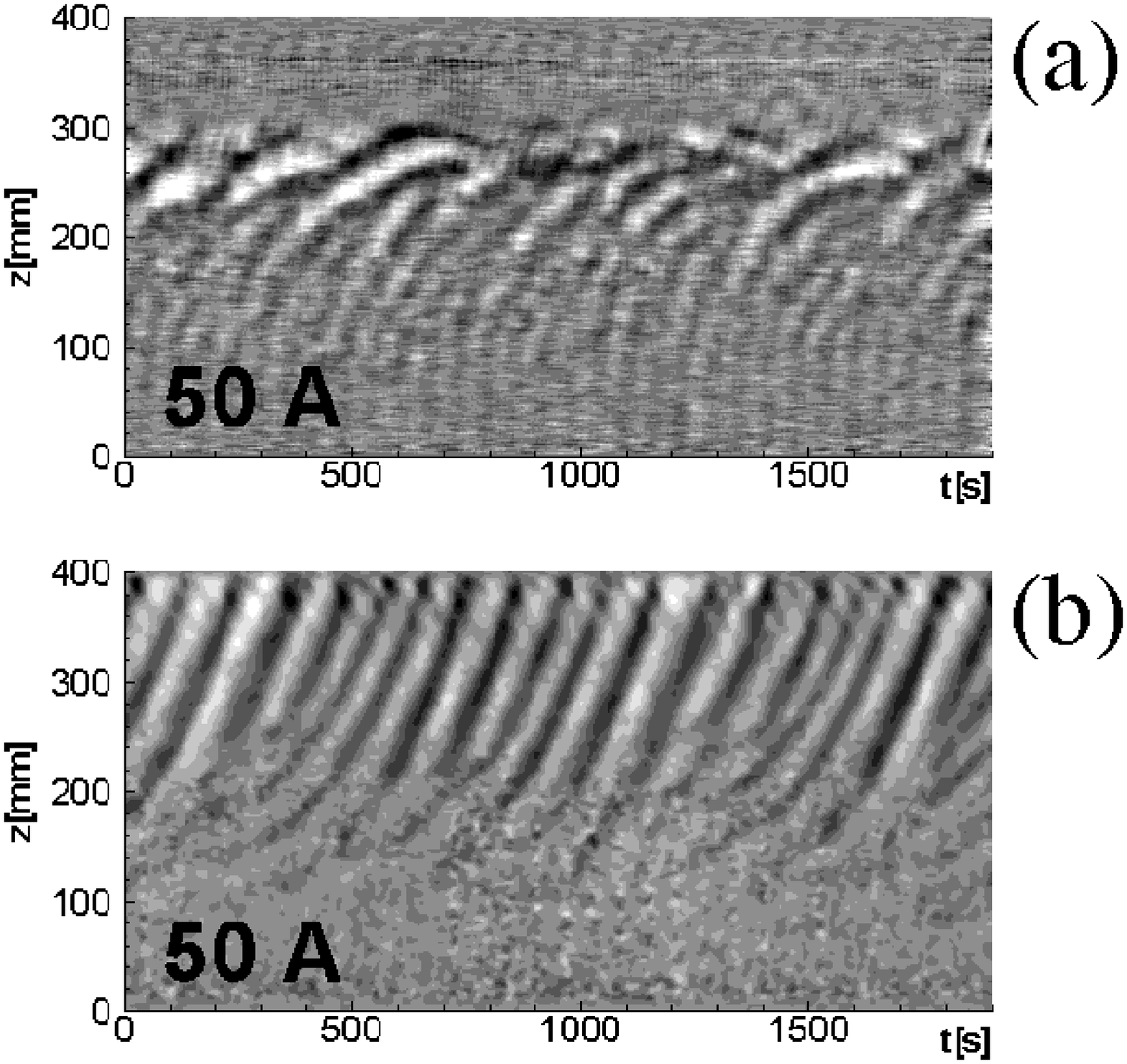}
\caption{The measured axial velocity perturbation for $Re=2975$ and $I_{coil}=50$ A, showing
the appearance of MRI as a travelling wave.
(a) PROMISE 1 experiment with $I_{rod}=6000$ A. (b) PROMISE 2 experiment with $I_{rod}=7000$ A.}
\end{center}
\end{minipage}
\end{vchfigure}

For all those reasons, some changes of the 
axial boundary conditions have been implemented in a modified experiment 
which is called now PROMISE 2.
As suggested by a thorough analysis of Szklarski \cite{ANJACEK2}, the 
Ekman pumping can be minimized by using split rings, the inner one rotating with 
the inner cylinder and the outer rotating with the outer cylinder,
with a splitting position at  0.4 of the gap width (Figure 25). Both upper and lower 
rings are now made of insulating material. 
Since the ultrasonic transducers are rotating with the outer
ring (i.e. with the slow rotation rate of the outer cylinder) their signal must
be transmitted by slip rings to the computer.

Without going into the detailed results of the PROMISE 2 experiment, which 
will be published elsewhere, we can state
that these modifications bring about a drastic improvement of the
MRI wave  and a significant
sharpening of the transitions between stable and unstable regimes.
In Figure 26 we see that for PROMISE 2 the MRI wave goes through the entire 
volume
while it was stopped at the radial jet position in PROMISE 1.

\section{Conclusions and prospects}

The last ten years have seen tremendous progress in liquid 
metal experiments on the origin and the action of cosmic 
magnetic fields.
With the success of the
complementary sodium experiments in Riga and Karlsruhe in 1999 
it was shown that self-excitation works not only in computer 
programs on the basis of rather smooth flows, but also in 
real-world turbulent flows.
Kinematic dynamo theory has been shown to be robust
with respect to low levels of turbulence and complicated boundary
and interface conditions. The saturation mechanism in the Riga experiment 
is non-trivial as it results not only from a global pressure
increase but also from a significant redistribution of the flow.

While the Riga and Karlsruhe dynamos are
characterized by an amazing predictability, in some respect
one can learn even more from the efforts to make the VKS dynamo 
(and also the MDX dynamo) running.
After some modifications of the original concept, the VKS 
experiment was eventually successful in self-exciting a magnetic field. 
The fact that the observed eigenfield in VKS is essentially an
axisymmetric dipole, in contrast to the original prediction of an 
equatorial dipole, is an inspiring challenge to understand better the 
intricate field amplification loop the dynamo mechanism relies on.
The ''little brother'' of the VKS experiment, the VKG experiment in Lyon,
has yielded fascinating field reversals when complemented by an external 
amplification loop that mimics a sort of $\alpha$ effect.

On the way to the final rotating torus experiment, the Perm group
has obtained important results concerning the mean-field coefficients in turbulent 
flows. 

After dynamo action has thus been proven, one observes presently some
tendency to take a breath and turn back to somewhat smaller machines 
in order to study MHD instabilities and wave phenomena. The identification 
of the MRI is only one aspect in this direction, though an important one 
due to the enormous astrophysical implications of this instability.
Apart from standard MRI (Maryland, Princeton, New Mexico) 
and helical MRI (Dresden-Rossendorf) there are other
magnetic instabilities that are capable of destabilizing hydrodynamically 
stable flows or even fluids at rest. Among them we have to note 
the Taylor-Vandakurov instability (with a current flowing through the 
liquid) \cite{VANDAKUROV,TAYLER,SPRUIT}
and the ''azimuthal MRI'' (AMRI) \cite{RUEDIGERHOLLERBACHSCHULTZ} based on an
purely azimuthal field. Future experiments on those instabilities are
very desireable.

There are many wave phenomena in rotating fluids under the influence
of (externally applied or self-excited) magnetic fields
which still  deserve a deeper understanding. 
At this point we observe a revival of the  activities in the sixties
on  Alfv\'{e}n wave studies with liquid metals.
In this respect, it might also be interesting that presently
magnetic fields are available \cite{HOCHFELD} 
which are so strong that for potassium and sodium
the Alfv\'{e}n velocity exceeds the sound velocity.
The small scale experiments in Maryland and Grenoble have provided 
a wealth of data, and the 3 m sphere in Maryland
will be a fascinating tool for extending those investigations into 
extreme parameter regions.

An interesting direction of future research could be the 
experimental investigation of precession driven dynamos, for which
a water test experiment had been carried out by L\'{e}orat et al. \cite{LEORAT1,LEORAT2}.
Numerical work by Tilgner \cite{TILGNERPRECESSION} points
to a critical $Rm$ of around 200 which makes a laboratory 
experiments not unrealistic. Of course, much more 
optimization would be necessary before such an experiment could be designed.

If one had a wish for free, one could  think of constructing
a huge self-sustaining nonlinear dynamo \cite{RINCON} 
(or ''self-creating dynamo'' \cite{RAEDLERFUCHS}).
One could start, for example, in the hydrodynamic stable regime of a
Taylor-Couette flow (mimicking
a Keplerian flow) which could be destabilized, via MRI, by an externally
applied magnetic field. Suppose now that the resulting flow
would act as a dynamo, the resulting 
eigenfield could possibly replace the initially applied magnetic field as
a trigger for the instability.

Having another wish for free, one could also think about
the experimental realization of a fluctuation dynamo, based on more 
or less homogeneous isotropic turbulence \cite{SCHEKOCHIHIN}.
For those flows, critical $Rm$ around 200 have been found in recent simulations.   %!!!
However, simple Kolmogorov scaling arguments tell 
us that such a ''James-Bond-dynamo'' 
(''...shaken, not stirred...'') is only possible with a giant 
input power of many Megawatts.

In spite of the fact that  liquid metal experiments should not 
be expected to be 
perfect Bonsai models of any real astrophysical systems, 
experimental work has already started to change some views on those 
natural systems. 
Maybe that the Riga dynamo will once
become a model of the double helix nebula close to the galactic 
center \cite{MORRIS,SHUKUROVNEBULA}, 
maybe that the partitioning  of
$\alpha$ and $\Omega$ effects in the VKS dynamo will 
re-animate the old Babcock-Leighton theory of the
solar dynamo \cite{BABCOCK}, maybe that  the PROMISE experiment 
will illuminate
the possible  action of helical MRI in cold parts of accretion disks 
were standard MRI cannot work.

\begin{acknowledgement}
Support from Deutsche Forschungsgemeinschaft in the framework of SFB 609,
from European Commission under contract 028679, and from Leibniz-Gemeinschaft
in the framework of the SAW project PROMISE is gratefully acknowledged. 
We are indebted to Raul Avalos-Zu\~{niga}, Michael Christen, 
Andr\'{e} Giesecke, Thomas Gundrum, Uwe G\"unther, Rainer Hollerbach, 
Sa\v{s}a Kenjere\v{s}, Jacques L\'{eorat},
Olgerts Lielausis, Ernests Platacis, J\={a}nis Priede, 
Karl-Heinz R\"adler, G\"unther R\"udiger,
Luca Sorriso-Valvo, Jacek Szklarski, and Mingtian Xu
for the fruitful collaboration over the course of many years.

\end{acknowledgement}

% ---- Bibliography ----
%
%\begin{references}
%\begin{thebibliography}{200}

%\end{references}

\end{document}